\documentstyle[psfig,graphicx,times]{mn}

\newif\ifAMStwofonts

\def\etal{et al.\ }
\def\eg{{ e.g.\ }}

\def\ie{{ i.e.\ }}

\def\approxlt{{\stackrel{\raisebox{-1pt}{{$\scriptscriptstyle<$}}}
{{\scriptscriptstyle \sim}} }}

\def\arc{{\rm\thinspace arcsec}}
\def\cm{{\rm\thinspace cm}}

\def\deg{$^\circ$}
\def\erg{{\rm\thinspace erg}}

\def\ha{H$\alpha$}
\def\hb{H$\beta$}

\def\keV{{\rm\thinspace keV}}
\def\m{{\rm\thinspace m}}

\def\km{{\rm\thinspace km}}
\def\kpc{{\rm\thinspace kpc}}

\def\Mpc{{\rm\thinspace Mpc}}
\def\Msun{\hbox{$\rm\thinspace M_{\odot}$}}

\def\s{{\rm\thinspace s}}
\def\yr{{\rm\thinspace yr}}

\def\apc{\rm atom cm$^{-2}$}

\def\ergpspcmsq{\hbox{$\erg\cm^{-2}\s^{-1}\,$}}

\def\ergpspcmsqpA{\hbox{$\erg\s^{-1}\cm^{-2}$\AA$^{-1}\,$}}

\def\ergps{\hbox{$\erg\s^{-1}\,$}}

\def\kmps{\hbox{$\km\s^{-1}\,$}}

\def\Msunpyr{\hbox{$\Msun\yr^{-1}\,$}}

\def\pcmsq{\hbox{$\cm^{-2}\,$}}

\def\psqcm{\hbox{$\cm^{-2}\,$}}

\def\kmpspMpc{\hbox{$\kmps\Mpc^{-1}$}}

\def\oid{[OI]$\lambda\lambda 6300,6363$\thinspace\AA{}}
\def\niid{[NII]$\lambda\lambda 6548,6584$\thinspace\AA{}}
\def\siid{[SII]$\lambda\lambda 6717,6731$\thinspace\AA{}}

\ifoldfss
  \ifCUPmtlplainloaded \else
    \NewTextAlphabet{textbfit} {cmbxti10} {}
    \NewTextAlphabet{textbfss} {cmssbx10} {}
    \NewMathAlphabet{mathbfit} {cmbxti10} {} 
    \NewMathAlphabet{mathbfss} {cmssbx10} {} 
  \fi
  \ifAMStwofonts
    \ifCUPmtlplainloaded \else
      \NewSymbolFont{upmath} {eurm10}
      \NewSymbolFont{AMSa} {msam10}
      \NewMathSymbol{\upi}     {0}{upmath}{19}
      \NewMathSymbol{\umu}     {0}{upmath}{16}
      \NewMathSymbol{\upartial}{0}{upmath}{40}
      \NewMathSymbol{\leqslant}{3}{AMSa}{36}
      \NewMathSymbol{\geqslant}{3}{AMSa}{3E}

      \let\leq=\leqslant 
       
    \fi
  \fi
\fi 

\ifnfssone
  \newmathalphabet{\mathit}
  \addtoversion{normal}{\mathit}{cmr}{m}{it}
  \addtoversion{bold}{\mathit}{cmr}{bx}{it}
  \newmathalphabet{\mathbfit} 
  \addtoversion{normal}{\mathbfit}{cmr}{bx}{it}
  \addtoversion{bold}{\mathbfit}{cmr}{bx}{it}
  \newmathalphabet{\mathbfss} 
  \addtoversion{normal}{\mathbfss}{cmss}{bx}{n}
  \addtoversion{bold}{\mathbfss}{cmss}{bx}{n}
  \ifAMStwofonts
    \ifCUPmtlplainloaded \else
      \UseAMStwoboldmath
      \makeatletter
      \new@mathgroup\upmath@group
      \define@mathgroup\mv@normal\upmath@group{eur}{m}{n}
      \define@mathgroup\mv@bold\upmath@group{eur}{b}{n}
      \edef\UPM{\hexnumber\upmath@group}
      \new@mathgroup\amsa@group
      \define@mathgroup\mv@normal\amsa@group{msa}{m}{n}
      \define@mathgroup\mv@bold\amsa@group{msa}{m}{n}
      \edef\AMSa{\hexnumber\amsa@group}
      \makeatother
      \mathchardef\upi="0\UPM19
      \mathchardef\umu="0\UPM16
      \mathchardef\upartial="0\UPM40
      \mathchardef\leqslant="3\AMSa36
      \mathchardef\geqslant="3\AMSa3E

      \let\leq=\leqslant 

    \fi
  \fi
\fi 

\ifnfsstwo
  \DeclareMathAlphabet{\mathbfit}{OT1}{cmr}{bx}{it}
  \SetMathAlphabet\mathbfit{bold}{OT1}{cmr}{bx}{it}
  \DeclareMathAlphabet{\mathbfss}{OT1}{cmss}{bx}{n}
  \SetMathAlphabet\mathbfss{bold}{OT1}{cmss}{bx}{n}
  \ifAMStwofonts
    \ifCUPmtlplainloaded \else
      \DeclareSymbolFont{UPM}{U}{eur}{m}{n}
      \SetSymbolFont{UPM}{bold}{U}{eur}{b}{n}
      \DeclareSymbolFont{AMSa}{U}{msa}{m}{n}
      \DeclareMathSymbol{\upi}{0}{UPM}{"19}
      \DeclareMathSymbol{\umu}{0}{UPM}{"16}
      \DeclareMathSymbol{\upartial}{0}{UPM}{"40}
      \DeclareMathSymbol{\leqslant}{3}{AMSa}{"36}
      \DeclareMathSymbol{\geqslant}{3}{AMSa}{"3E}

      \let\leq=\leqslant 

    \fi
  \fi
\fi 

\ifCUPmtlplainloaded \else
  \ifAMStwofonts \else 
    \def\upi{\pi}
    \def\umu{\mu}
    \def\upartial{\partial}
  \fi
\fi

\title{The peculiar cooling flow cluster RX J0820.9+0752}
\author[Bayer-Kim \etal]
       {C. M. Bayer-Kim$^1$, C. S. Crawford$^1$, S. W. Allen$^1$, A. C. 
Edge$^2$ and
        A. C. Fabian$^1$ \\
         $^1$Institute of Astronomy, Madingley Road, Cambridge CB3 0HA\\
         $^2$Department of Physics, University of Durham, South Road,
Durham  DH1 3LE }

\date{Submitted 2002: June}

\pubyear{2002}

\begin{document}

\maketitle

\label{firstpage}

\begin{abstract}

\noindent We present observations of the cluster of galaxies
associated with the X-ray source RX~J0820.9+0752 and its
dramatic central cluster galaxy in the optical and X-ray wavebands.
Unlike other cooling flow central cluster galaxies studied in detail,
this system does not contain a powerful radio source at its core, and
so provides us with an important example where we expect to see only
the processes
directly due to the cooling flow itself. A 9.4\thinspace ks Chandra
observation shows that the hot intracluster gas is
cooling within a radius of 20\kpc\ at a rate of a few tens of solar
masses per year. The temperature profile is typical of a cooling flow
cluster and drops to below 1.8\keV\ in the core. Optical images taken with
the AAT and HST show that the central galaxy is
embedded in a luminous (L$_{H\alpha}\sim 5\times 10^{42}$\ergps),
extended line-emitting nebula that coincides spatially with a bright
excess of X-ray emission, and separate,
off-nucleus clumps of blue continuum that form part of a patchy
structure arcing away from the main galaxy. The X-ray/H$\alpha$
feature is reminiscent of the 40\kpc\ long filament observed in A 1795
which is suggested to be a cooling wake, produced
by the motion of the central cluster galaxy through the intracluster
medium. We present optical spectra of the central cluster galaxy and its
surroundings, and find that the continuum 
blobs show stronger line emission, differing kinematic properties
and more extreme ionization ratios than the surrounding nebula.
Accounting for the strong intrinsic reddening and its significant
variation over the extent of the line emitting region, we have fit the
continuum spectra of the blobs and the nucleus using empirical stellar spectra
from a library. We found that continuum emission from early main
sequence stars can account for the blue excess light in the
blobs. Kinematical properties associate the gas in the system with a
nearby secondary galaxy, suggesting some kind of tidal interaction
between the two. We suggest that the secondary galaxy
has moved through the cooling wake produced by the central cluster
galaxy, dragging some of the gas out of the wake and triggering the
starbursts found in the blobs.
\end{abstract}


\section{Introduction}
\label{introduction}

The hot intracluster medium in the cores of many clusters of galaxies
emits so much energy in the X-ray band that the cooling time is
substantially shorter than the cluster age. The resulting decrease
in gas pressure leads to a highly subsonic
inflow of material towards the cluster centre -- a cooling flow
(Fabian 1994). Recent X-ray spectra from {\sl XMM-Newton} have
confirmed the short central cooling times and low central cluster
temperatures required by a cooling flow, but show a surprising deficit
of spectral lines from gas cooling below $1-2$\keV{} (Peterson \etal 2001;
Kaastra \etal 2001). The rates of
cooling (or `mass deposition rate') thus appear to be reduced from
those deduced from earlier X-ray missions, and in the `classical' (\ie
pre-{\sl XMM-Newton} and {\sl Chandra}) picture of cooling flows. It
is possible that the age of the cooling flow can be substantially
reduced to around a few Gyr, for example if the intracluster medium is
stirred up by infall of subclusters (Allen, Ettori \& Fabian
2001a). The current
data are only consistent with the previously-deduced high mass
deposition rates if excess intrinsic absorption is introduced to
selectively remove the emission below 1\keV\ (David \etal 2001;
Schmidt, Allen \& Fabian 2001; Ettori \etal 2001).

Alternatively, a possible sink for the `missing' soft X-ray luminosity
expected from
steady cooling flows could be the luminous (and excess) line emission
seen in the ultraviolet/optical waveband (Heckman \etal 1989; Crawford
\etal 1999), and dust emission in the sub-mm/far-infrared (Edge \etal
1999; Allen \etal 2001c) seen
associated with many central cluster galaxies in cooling flows. In a cold
mixing model, the gas below $1-2$\keV\ drops from the flow to
be rapidly cooled by being churned with embedded cold gas clouds, and
then the `missing' energy is re-emitted at longer wavelengths
(Fabian \etal 2002).

Another explanation for the observed
properties has recently been put foward by Voigt~\etal~(2002). They
suggest that cooling can be balanced by heat conduction along
the temperature gradient in the outer
parts of a cluster with only a small cooling flow operating at the centre
($r\approxlt 20$\kpc), thereby reducing the mass deposition rates to
values consistent with observations. 

Most cooling flow models imply some correspondence in both the spatial and
luminous properties of the cooling gas seen in X-rays and the emission
seen in the lower-energy wavebands -- perhaps such as the X-ray/\ha\
filament extending from the central cluster galaxy in the A1795
cluster (Fabian \etal 2001a; Crawford \etal 2002 in
preparation), or the associated X-ray/emission-line regions seen in
objects such as A2199 (Johnstone \etal 2002), Virgo (Young, Wilson \& Mundell
2002), or A2390 (Allen \etal 2001a). These
line-luminous central cluster galaxies, however, are also sites for
extended regions of massive star formation that are presumed to be a
final sink for material cooling from the intracluster medium
(e.g. Crawford \& Fabian 1993).
Ionization from these stars may also be a large contribution to
powering the emission-line nebulae.

A further heat source which could balance radiative cooling in a
cluster is a central
radio source. Most central cluster galaxies have a radio source with jets
supplying energy into the surrounding gas, often blowing bubbles of
relativistic plasma (e.g. McNamara \etal 2000; Fabian \etal 2000). At
issue here is whether these bubbles lead to local heating of the
intracluster medium (Churazov~\etal~2002; Churazov \etal 2001;
Br\"uggen \& Kaiser 2001; Reynolds, Heinz \& Begelman 2002) or
buoyantly rise and transport the energy to
large radii (Fabian \etal 2002). It is also possible that cooling is
balanced by a combination of heating by the radio source and
conduction (Ruszkowski \& Begelman 2002). In this paper we present a central
cluster galaxy which has a particularly weak central radio source. 
(Of course such sources may go through cycles of activity and our
object may
just be in a low state.) It also appears to have a strong correspondence
between its cooling
X-ray emission and its properties at optical wavelengths. The galaxy
is at the centre of the cluster associated with the {\sl ROSAT} X-ray source
RX~J0820.9+0752\footnote{For convenience, we shall use `RX~J0821'
throughout the rest of this paper to refer to the central galaxy and its
associated line-emitting nebula.}, the coordinates of the nucleus of
the CCG being $\alpha=08$:21:02.28, $\delta=+07$:51:46.8 (J2000). The
cluster was discovered as part
of the extended Brightest Cluster Sample (Ebeling \etal 2000), and
optical spectroscopy on the central galaxy yielded a
redshift of 0.110 (Crawford \etal 1995). It has a {\sl ROSAT}
$0.1-2.4$\keV\ (rest frame) luminosity of
L$_X=2.09\times10^{44}$\ergps\footnote{We assume a cosmology with a
Hubble constant of $H_0=50$\kmpspMpc\ and a cosmological deceleration
parameter of $q_0=0.5$ throughout this paper.}, and an estimated
temperature of 4.4\keV. The spectrum of the central cluster galaxy
showed it to have strong \ha+[NII] emission lines (a slit \ha\
luminosity of around $3.3\pm0.2\times10^{41}$\ergps) but with a lack of
\hb\ that implied large amounts of intrinsic obscuration to be present
in this system (Crawford \etal 1995).

RX~J0820.9+0752 is a strong CO emitter (Edge 2001) with an implied molecular
gas mass of 3.9$\pm0.4\times 10^{10}$~M$_\odot$ and offset in
velocity from the main galaxy by about $+260$\kmps. The CO emission is
centred on the central galaxy but extended to the W by about 5\arc{}
(Edge \& Frayer 2002 in prep.). RX~J0820.9+0752 is also
detected by {\it IRAS} at 60$\mu$m (Edge 2001) implying a
dust mass of 2.2$\times 10^7$~M$_\odot$ for a dust temperature of 40~K.
A near-infrared
spectrum revealed Pa$\alpha$ extended over 2.4 \arc{} to the north of
the central galaxy, but there was no significant detection of any
\mbox{1-0~S~series} H$_2$ line emission, despite the strong CO
detection (Edge \etal 2002). 

\section{{\sl Chandra} X-ray observation}
\label{xrayobs}

The {\sl Chandra} observation of RX~J0820.9+0752 was carried out on 2000
October 6, using the Advanced CCD Imaging Spectrometer (ACIS). The
target was observed in the back-illuminated S3 detector and positioned
near the centre of node-1 on CCD 7. The net exposure time was
9.4\thinspace ks
and the focal plane temperature at the time of the observations was
$-120$C. We used the CIAO software and the level-2 events files
provided by the standard {\sl Chandra} pipeline processing for our
analysis.  Only those X-ray events with grade classifications of
0, 2, 3, 4 and 6 were used.

\begin{figure}
\vspace{0.5cm}
\hspace{0.8cm}\psfig{figure=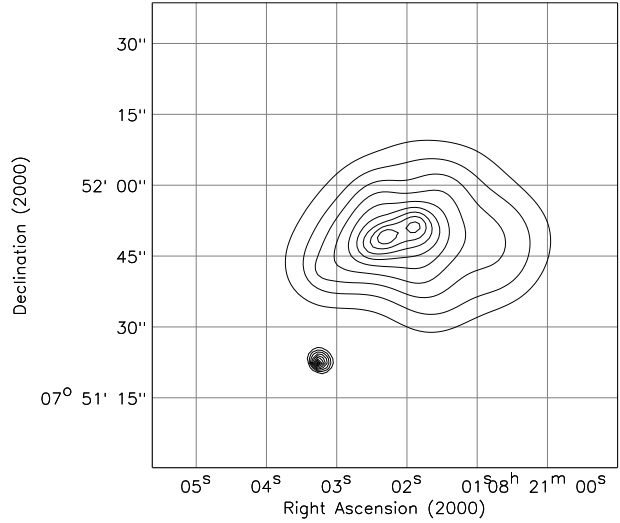,width=0.45 \textwidth,angle=0}
\caption{
\label{fig:xpic}
Contour plot of the $0.3-7.0$ keV {\sl Chandra} image of RX~J0820.9+0752,
adaptively smoothed using the code of Ebeling, Rangarajan \& White
(2002), with a threshold value of $3.5 \sigma$. The contours have
equal logarithmic spacing. }
\end{figure}

An immediate inspection of the {\sl Chandra} image
(Figures~\ref{fig:xpic} and \ref{fig:allpic}) shows that the cluster
emission appears extended: the bulk of the emission is centred on the
position of the central cluster galaxy, but there is a strong bright
extension over 8 \arc{} towards the NW ($\sim22$\kpc; 1 \arc{}
corresponds to a distance of $\sim$2.7\kpc\ at the redshift of
RX~J0820.9+0752). The 0.5-7\keV\ luminosity of the cluster emission is
$1.6\pm0.1\times10^{44}$\ergps, with a bolometric luminosity of
$2.5\times10^{44}$\ergps.

We also extracted separate spectra from two circular regions of 5
\arc{} radius -- one centred on the central galaxy, and a second
offset to the NW over the bulk of the extended emission. There were
too few photons in each spectrum to do precise modelling, but we found
upper limits to the intrinsic absorption column density of a few by
10$^{21}$\pcmsq.

\subsection{Spectral analysis}

Spectra were extracted from two circular annuli, centred on the X-ray
peak (at $\alpha=08$:21:02.4, $\delta=+07$:51:48; J2000), and spanning radii of
$0-50$ kpc (18.7 \arc) and $50-200$ kpc (75.3 \arc), respectively.
The emission from point sources was masked and excluded. The spectra
were grouped before fitting to contain a minimum of 20 counts per
pulse invariant channel, allowing $\chi^2$ statistics to be used. A
background spectrum was extracted from a source-free region of node-3
on the S3 chip. For the $0-50$ kpc annulus, data covering the
$0.6-7.0$ keV energy range were used. For the $50-200$ kpc annulus,
for which the background is more significant, our analysis was limited
to the $0.6-5.0$ keV band. Separate photon-weighted response matrices
and effective area files were constructed for each region using the
2001 August release of the ACIS-S calibration and response files.

The spectra have been analysed using the XSPEC code (version 11.01;
Arnaud 1996) and MEKAL plasma emission models of Kaastra \& Mewe
(1993; incorporating the Fe L calculations of Liedhal, Osterheld \&
Goldstein 1995). We have fit the spectra using a single-temperature
model, with a temperature ($kT$), metallicity ($Z$; measured relative
to the solar photospheric values of Anders \& Grevesse 1989, with the
various elements assumed to be present in their solar ratios) and
absorbed by a column density ($N_{\rm H}$) of cold gas
(Balucinska-Church \& McCammon 1992). The results of the fits are summarized in
Table 1. The (projected) temperature rises from a value of
$\sim 1.8$ keV within the central 50 kpc radius, to a value of $\sim
3$ keV between radii of $50-200$ kpc. This type of temperature profile
is typical of cooling flow clusters (e.g. Allen, Schmidt \& Fabian
2001b).

\begin{table}
\begin{center}
\caption{The results from the analysis of the annular spectra.
Temperatures ($kT$) are in keV, metallicities ($Z$) in solar units, 
and absorbing column
densities ($N_{\rm H}$) in units of $10^{20}$\apc. The total $\chi^2$ values 
and number of degrees of freedom (DOF) in the fits are listed in column 5.  
Error bars are the $1\sigma$ ($\Delta \chi^2=1.0$) confidence limits on a single
interesting parameter.  }
\begin{tabular}{ c c c c c c }
&&&&&  \\                                                                                                                   \hline
Model A     & ~ &  $kT$                    &    $Z$                   & $N_{\rm H}$        &  $\chi^2$/DOF \\

\hline
$0-50$ kpc    & ~ & $1.82^{+0.17}_{-0.16}$  & $0.41^{+0.15}_{-0.12}$  & $3.2^{+2.5}_{-2.2}$ &  44.1/50   \\
$50-200$ kpc  & ~ & $3.02^{+0.57}_{-0.46}$  & $0.40^{+0.23}_{-0.15}$  & $5.2^{+2.3}_{-2.2}$ &  104/100   \\

&&&&& \\
\hline
\label{xmodel}
\end{tabular}
\end{center}
\end{table}

\subsection{Cooling in the cluster core}
\label{cooling}

\begin{figure}
\vspace{0.5cm}
\hbox{
\hspace{0.2cm}\psfig{figure=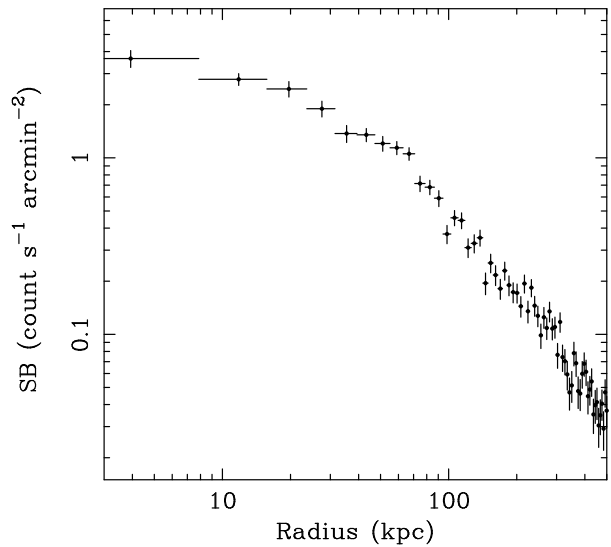,width=0.45 \textwidth,angle=270}}
\caption{The background-subtracted, flat-fielded, azimuthally-averaged 
radial surface brightness profile for RX~J0820.9+0752 in the $0.3-7.0$ keV 
band.}\label{fig:surbri}
\end{figure}

The azimuthally-averaged, $0.3-7.0$ keV X-ray surface brightness
profile for RX~J0820.9+0752 is shown in Figure~\ref{fig:surbri}. The profile
has been flat-fielded and background subtracted. The bin-size is 2
detector pixels (0.984 \arc).

\begin{figure}
\vspace{0.5cm}
\vbox{
\hspace{0.2cm}\psfig{figure=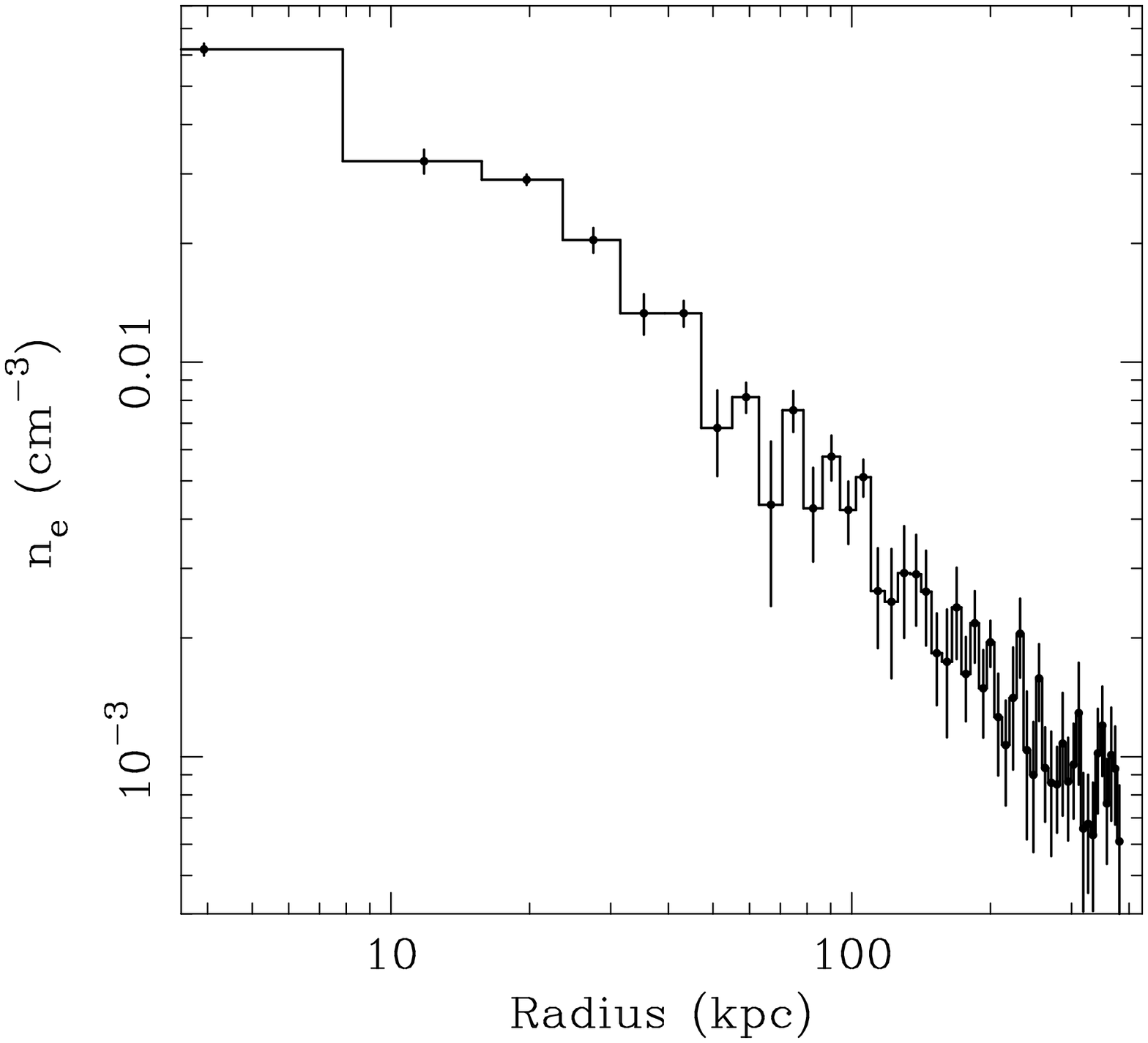,width=0.45\textwidth,angle=270}
\hspace{0.8cm}\psfig{figure=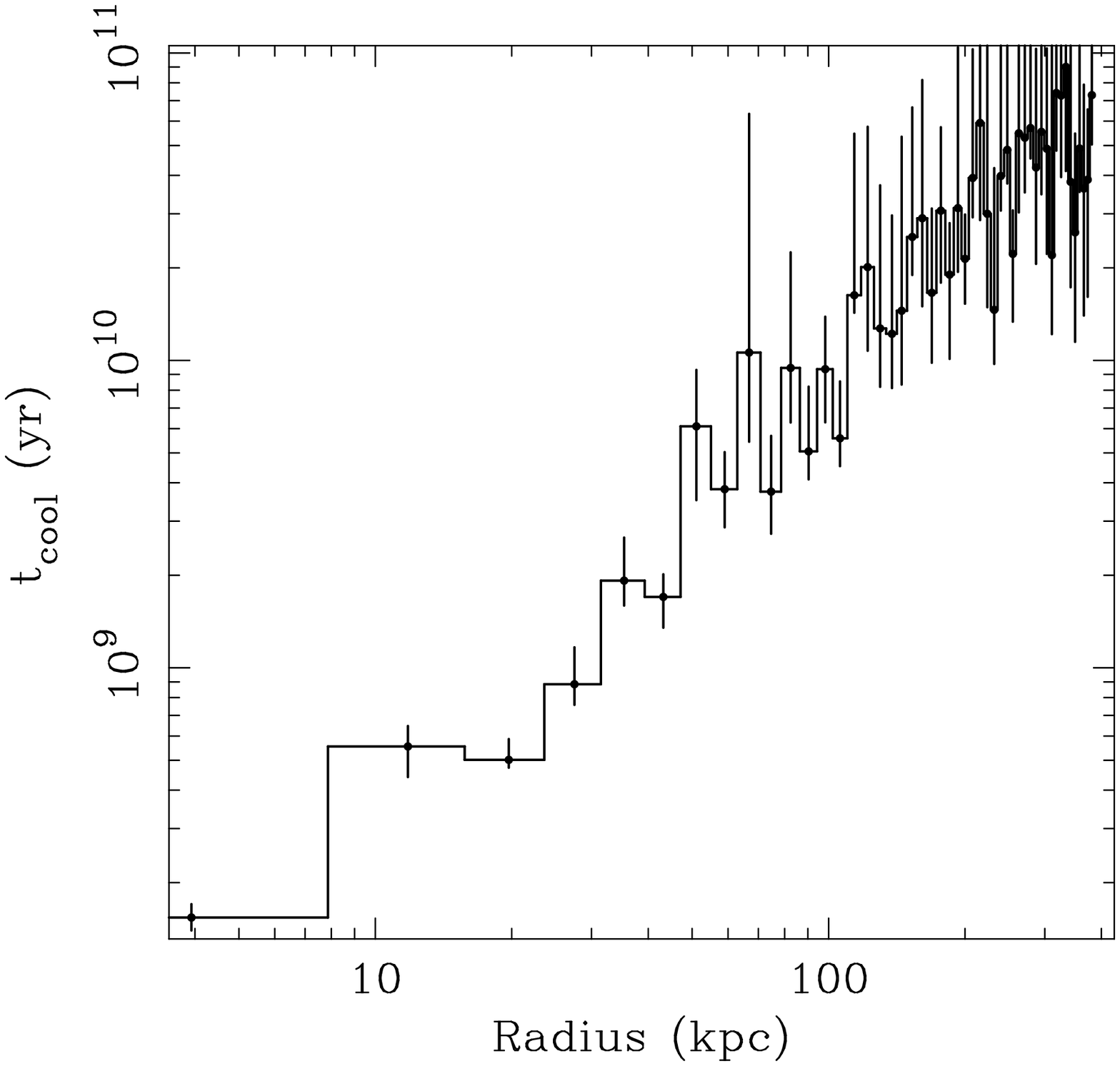,width=0.45\textwidth,angle=270}
\hspace{0.0cm}\psfig{figure=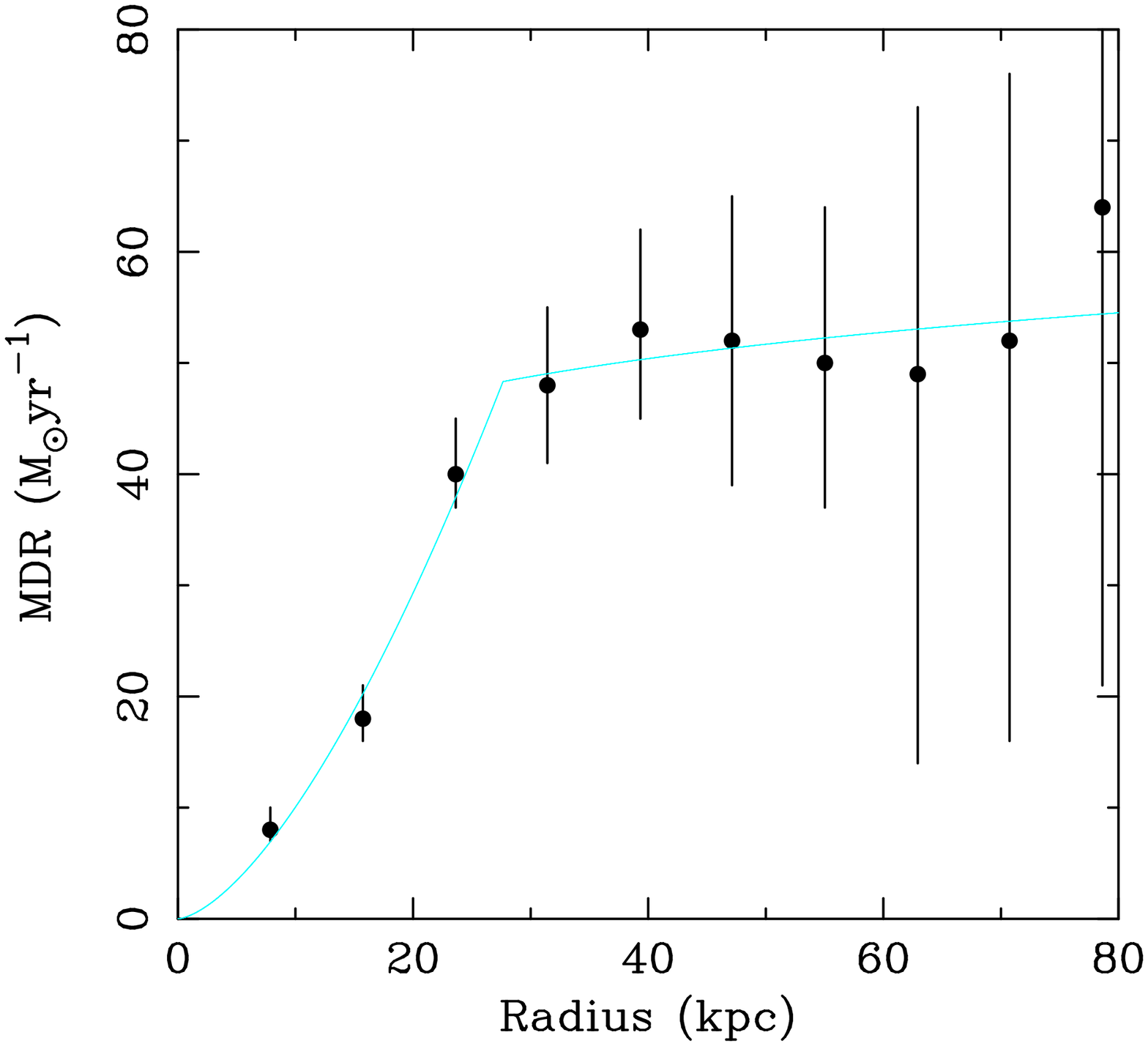,width=0.45 \textwidth,angle=270}
}
\vspace{0.1cm}
\caption{ 
\label{fig:xproj}
The results on (top) the electron density and (middle) the cooling
time,  determined from the X-ray image deprojection analysis. 
Error bars are the  $1\sigma$ errors determined from 100 Monte Carlo 
simulations. The lower panel shows the 
mass deposition rate (MDR) determined from  the image
deprojection analysis under the assumption that the central regions of
the cluster contain a steady-state, inhomogeneous cooling flow.  Error
bars are the 10 and 90 percentile  values from 100 Monte Carlo
simulations. The grey curve shows the best-fitting broken power-law
model.}
\end{figure}

We have carried out a deprojection analysis of the cluster, using the
observed temperature profile to constrain the analysis, as described
by e.g. Allen \etal (2001a). The inferred electron density, cooling
time and equivalent mass deposition rate (which parametrizes the
luminosity distribution in the cluster core, under the assumption that
the gas there is in a steady state cooling flow; \eg White, Jones \&
Forman 1997) are shown in Figure~\ref{fig:xproj}. When accounting for
projection effects, the temperature in the central 50 kpc region drops
to $1.67\pm0.12$ keV. The cooling time of the cluster gas is $<10^9$
yr within the central 20 kpc; the mass deposition rate within this
radius is $\sim 30$\Msunpyr (assuming that the gas there is in a
steady state cooling flow).

A cooling flow component (Johnstone et al. 1992) added to the
model for the spectrum of the X-ray flux from within 50~kpc gives an
improved fit over the single MEKAL model ($\Delta\chi^2=-7.2$). The mass
cooling rate obtained is $\dot M_{\rm spec}=47^{+11}_{-18}\Msunpyr$,
the metallicity $Z=0.8^{+0.39}_{-0.27}$, and intrinsic absorption
$N_{\rm H}=10\pm 0.4\times 10^{20}\psqcm$. The upper temperature of
the cooling flow component (and temperature of the additional MEKAL
component) is $kT=2.28^{+0.35}_{-0.30}\keV$. Whether there is excess
intrinsic absorption or not depends on the uncertain soft X-ray
response of ACIS-S (the data were fit only above 0.7~keV); $\dot
M_{\rm spec}$ drops to about $30\Msunpyr$ if $N_{\rm H}\sim 5\times
10^{20}\psqcm$.

\begin{figure}
\vbox{
\psfig{figure=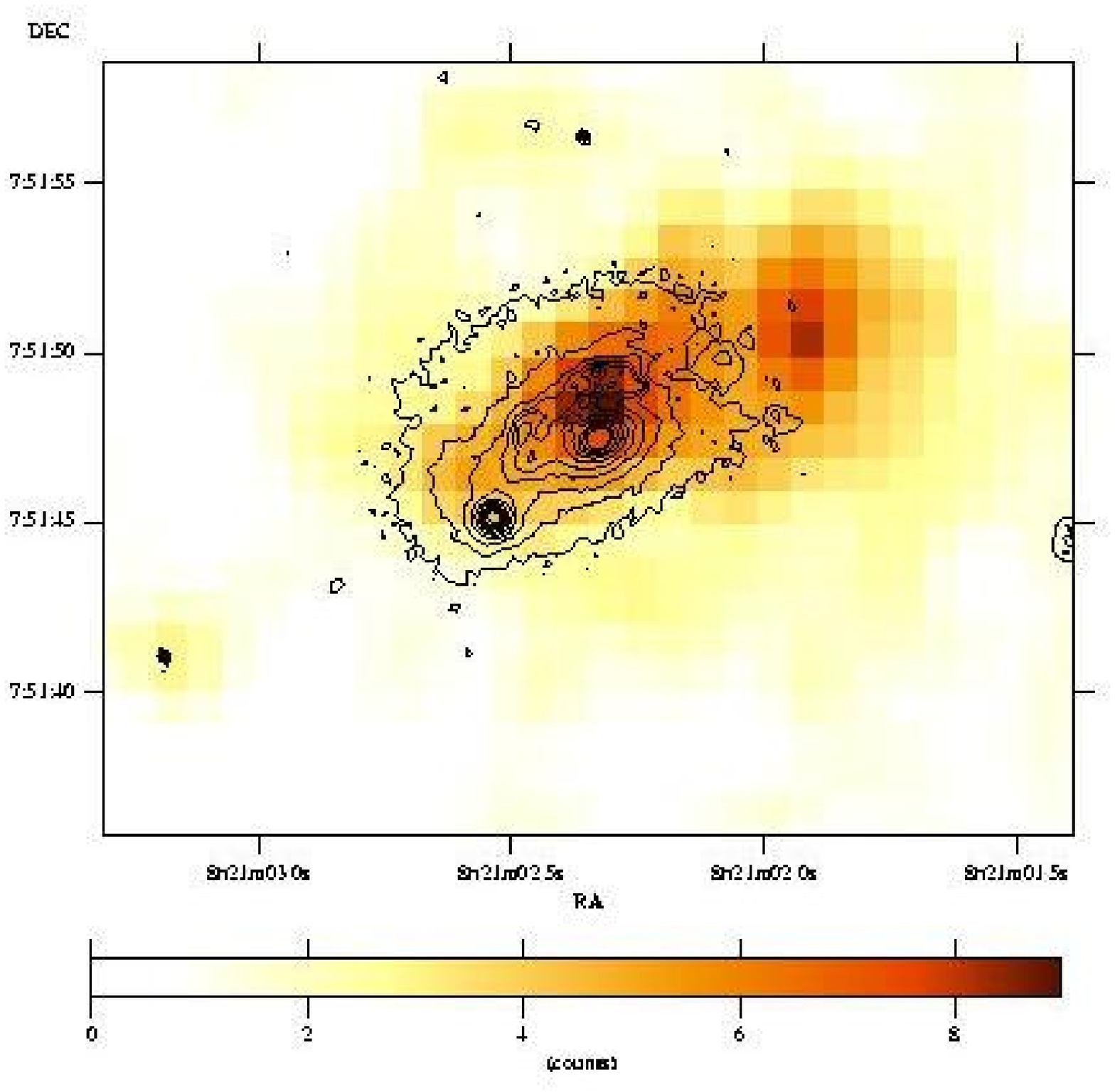,width=0.5\textwidth}
\psfig{figure=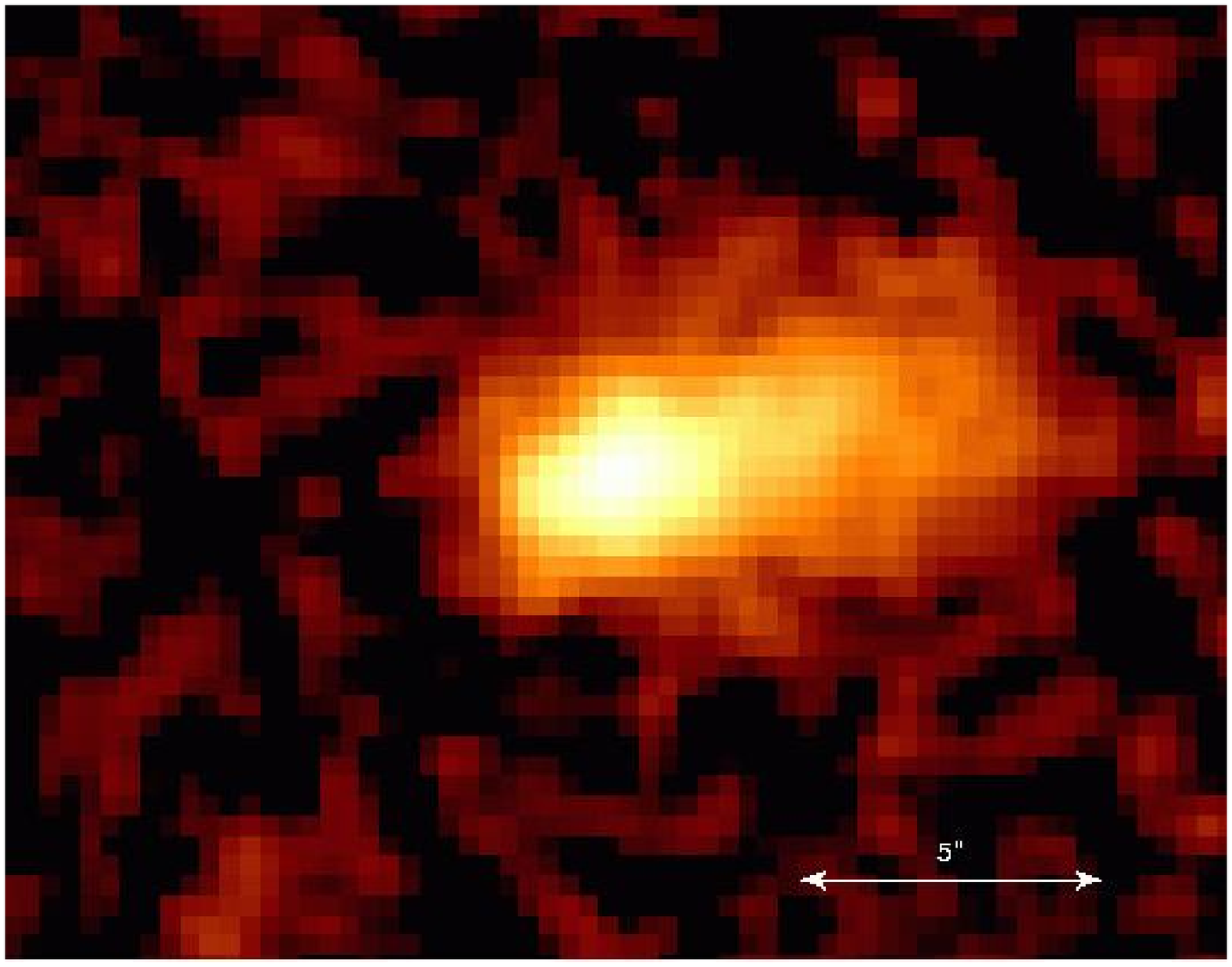,width=0.5\textwidth}
\psfig{figure=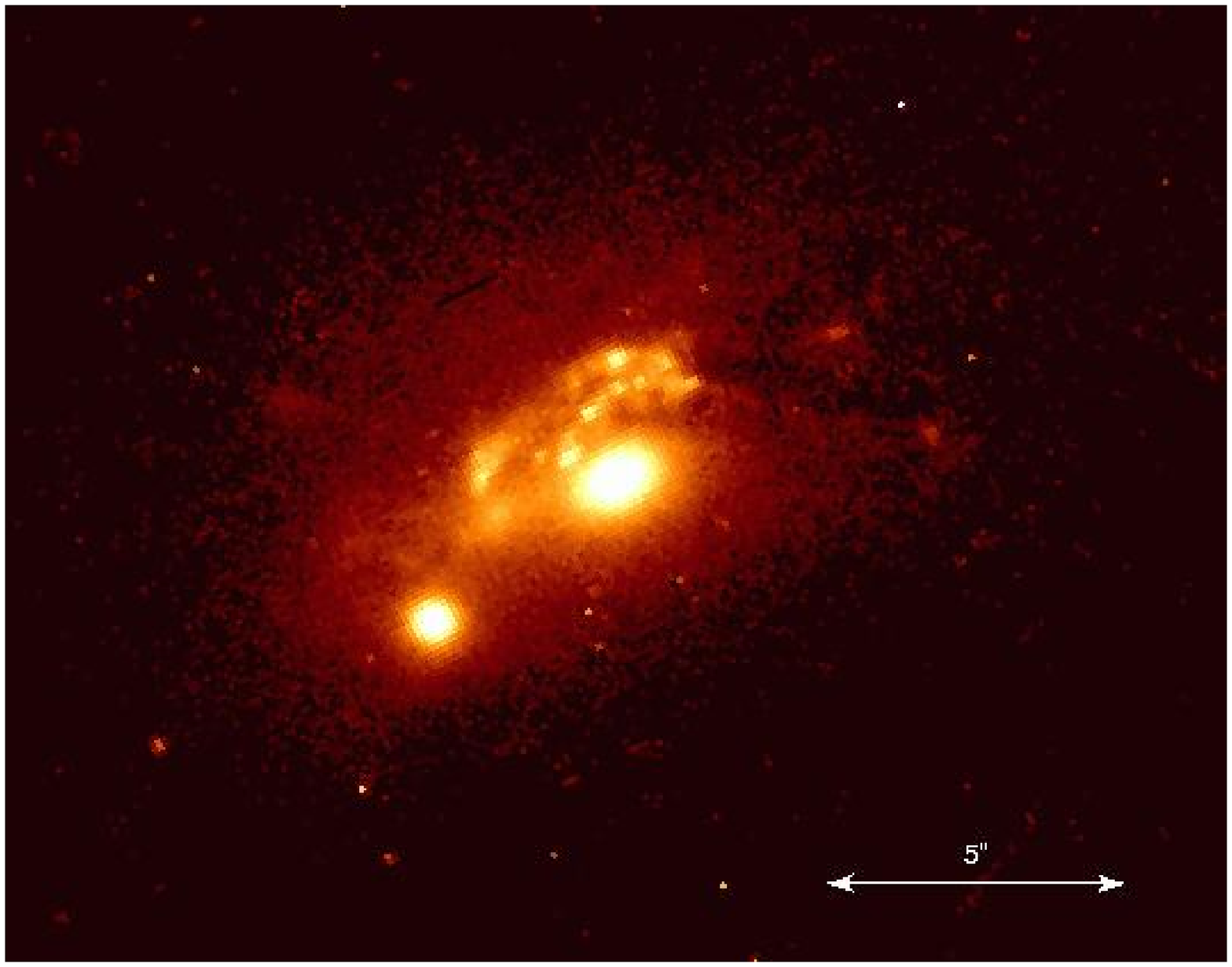,width=0.5\textwidth} }
\caption{\label{fig:allpic}
The slightly smoothed $0.5-2$\keV{} {\sl Chandra} image of the
cluster emission from RX~J0820.9+0752 with the optical contours
overlaid (top). The TTF image of the (continuum-subtracted)
\ha\ emission around the central galaxy in RX~J0820.9+0752 (middle). The image
shown is 20.5$\times$16\arc. The F606W HST image of the
central regions of RX~J0820.9+0752 on the same scale as the middle
image (bottom). North is to the top and East to the left in all the
figures.}
\end{figure}

\section{Radio observations}
\label{radio}

RX~J0821 is a relatively unusual central cluster galaxy in the 
radio. 
From a complete VLA snapshot campaign at 5~GHz (Edge~et al., in
preparation), RX~J0821 is the weakest of all the BCS central
cluster galaxies with detected optical line emission. The 
flux density of 0.73$\pm0.09$~mJy at 4.89~GHz coupled to the
2.26$\pm$0.14~mJy at 1.4~GHz in the First Survey (Becker \etal 1995)
give a canonical spectral index ($-0.90^{+0.16}_{-0.14}$) and
low radio power ($1.2\times 10^{23}$\thinspace WHz$^{-1}$ at 1.4~GHz).

The radio source shows no discernible structure on the $2-5$\arc{}
scale (our 5~GHz C-array and the First Survey data have comparable
resolution), although it lies to the NW ($\alpha=08$:21:02.2,
$\delta=+07$:51:48.78) of the core of the dominant galaxy.
More detailed VLA imaging is required to identify whether the radio
emission is directly related to the off nuclear features. (It
coincides with H$\alpha$SB defined in Sec. \ref{obsred}.)

As an aside the most striking feature in the radio maps is a strong
head-tail galaxy 30\arc{} to the south-west of the cluster core. The head of
the radio emission is coincident with a moderately bright elliptical
galaxy in the HST image and therefore this is very probably a cluster
member. The tail
of the radio emission extends to 35~\arc{} in length and has a
position angle approximately 35~degrees from the dust plume to the
west of the central galaxy. This head-tail source is suggestive of a
galaxy falling into the cluster from a western direction.

If the radio emission was only due to star formation, this would imply
an $\dot{M}_{\rm Star}=32$\Msunpyr (using the relation of Yun, Reddy
\& Condon 2001, corrected for the differing $H_0$), which
is in good agreement with the X-ray results.

\section{{\sl TTF\ } observations}
RX~J0821 was observed on 2000 April 15 using the Taurus tunable
filter (TTF; Bland-Hawthorn \& Jones 1998) on the Anglo-Australian
telescope. The TTF was
tuned to a bandwidth of 30\thinspace\AA, and a total exposure of
1800\thinspace s was taken
through a band centred at 7291\thinspace\AA\ to encompass redshifted \ha\
(with a small degree of contamination of [NII]). In order to estimate
the continuum level at \ha, two neighbouring bands at 7258\thinspace\AA\ and
7383\thinspace\AA\ were also observed in `straddle shuffle' mode, again for a
total exposure of 1800s. The resulting continuum-subtracted \ha\ image
(Figure~\ref{fig:allpic}) clearly shows that the \ha\ emission forms an
extended nebula stretching over $\sim7$\arc{} (19\kpc) to the NW.

\section{{\sl Hubble Space Telescope} observation}

A Hubble Space Telescope (HST) WFPC2 image of RX~J0821 was taken on
2000 February 12 for an exposure time of $2\times300$\thinspace s, 
using the F606W filter. The two images were combined using the
standard {\it IRAF STSDAS} routines to remove cosmic rays
and were then filtered for
hot and cold pixels. The F606W filter
corresponds to wide $V$, covering a wavelength
range of $\Delta\lambda=4485-5907$\thinspace\AA\ in the restframe of the
object so does not include [NII]+\ha. The WFPC
image is shown in Figure~\ref{fig:allpic}. 

The primary feature in the HST image is the central galaxy and its
smaller companion $\sim3.8$ \arc{} to the SE. Two very bright lines of
clumped emission arc from the NW over to the E, with several other
isolated blobs also apparent scattered further out to the NW and E.
The whole system is embedded in diffuse emission, although with a
sharp \lq bite' dimming this extended halo starting just beyond 3
arcseconds to the NW.

\section{Optical spectra}

\subsection{Observations and data reduction}
\label{obsred}
\begin{figure}
\psfig{figure=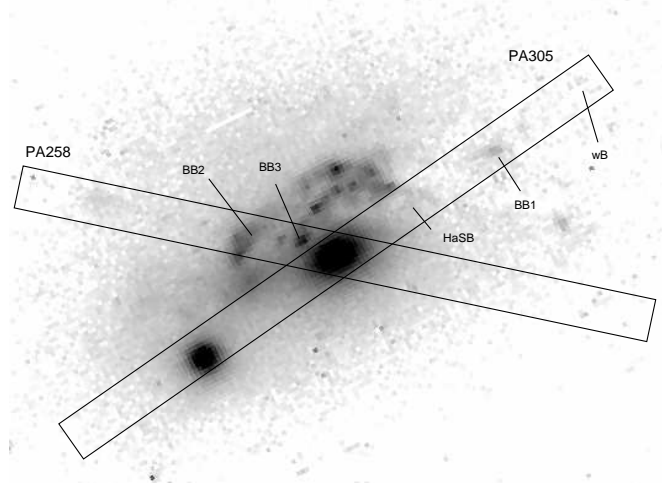,width=0.5\textwidth,angle=0}
\caption{\label{fig:slits} 
Schematic of the slit positions and the location of specific regions
of the RX~J0821 system referred to in the text, overlaid on the HST
image. }
\end{figure}

Optical spectra of RX~J0821 were taken
on 2000 January 3 using the 4.2\m{} William Herschel Telescope (WHT)
of the Roque de los Muchachos Observatory. The Intermediate dispersion
Spectroscopic and Imaging System (ISIS) was used to observe the object
at position angles of 305\deg and 258\deg (henceforth PA305 and PA258
respectively; see Figure~\ref{fig:slits}), using both the red and blue
arms of the instrument. The 5400\thinspace\AA{} dichroic was used to split the
data, which unfortunately starts diminishing the continuum light
noticeably beyond about 4600\thinspace\AA\ in the rest frame of the system. The
R158B grating was used on the blue arm, with the EEV12 chip as the
detector to give a dispersion of 1.62\thinspace\AA{}/pix centred at
about 4145\thinspace\AA{}. Similarly, using the R158R grating with
the TEK4 CCD on the red arm gave a dispersion of
2.9\thinspace\AA{}/pix centred at 6343\thinspace\AA. The spatial dispersion of
the data on the blue and red arms was 0.2 \arc/pix and 0.36
\arc/pix respectively. A one arcsecond slit was used throughout all
the observations, to match the seeing conditions of
$0.9-1.1$\thinspace \arc{}. The individual exposure times were
generally 1800 seconds, combining to give a total exposure time of
11,300 seconds and 10,800 seconds for the red and blue arm at PA305
respectively, and 1800 seconds each at PA258. The airmass varied
between approximately 1.1 and 1.3 during the observations.

Basic reduction of the data was performed using standard IRAF
procedures: bias level subtraction, cosmic ray removal and
flat-fielding. The spectra were
straightened to correct for any spatial dispersion due to atmospheric
refraction and co-added (where appropriate) to produce a final image
for each position angle and arm. Each co-added image was then
wavelength-calibrated and
sky-subtraction was performed by producing a sky frame from averaged
cross sections well off the actual galaxy from each spectrum. Flux
calibration was carried out using observations of two standard stars,
taken during the same night as the object spectra, with the
same wavelength coverage and instrumental setup. The flux calibration
between the two arms of ISIS looks convincing and gives a consistent
response function. The accuracy of the calibration is also reinforced
by the perfect match of the nuclear spectrum between the R and B arms.
Corrections for atmospheric extinction and Galactic reddening were
applied. Appropriate hydrogen columns derived from the data of Stark
\etal~(1992) were used to calculate the amount of Galactic reddening
as {\sl E(B-V)} $=0.04$ using the formula of Bohlin, Savage \& Drake
(1978). The spectra
were de-redshifted using $z=0.110$ (Crawford \etal 1995), and finally
the data were binned spatially to improve the signal-to-noise and to
obtain a spectrum of specific regions with distinct spectral
properties. Note that due to the longer exposure time, the
signal-to-noise in PA305 is significantly better than that in PA258.

\begin{figure*}
\begin{minipage}{175mm}
\includegraphics[width=70mm,height=80mm,angle=270]{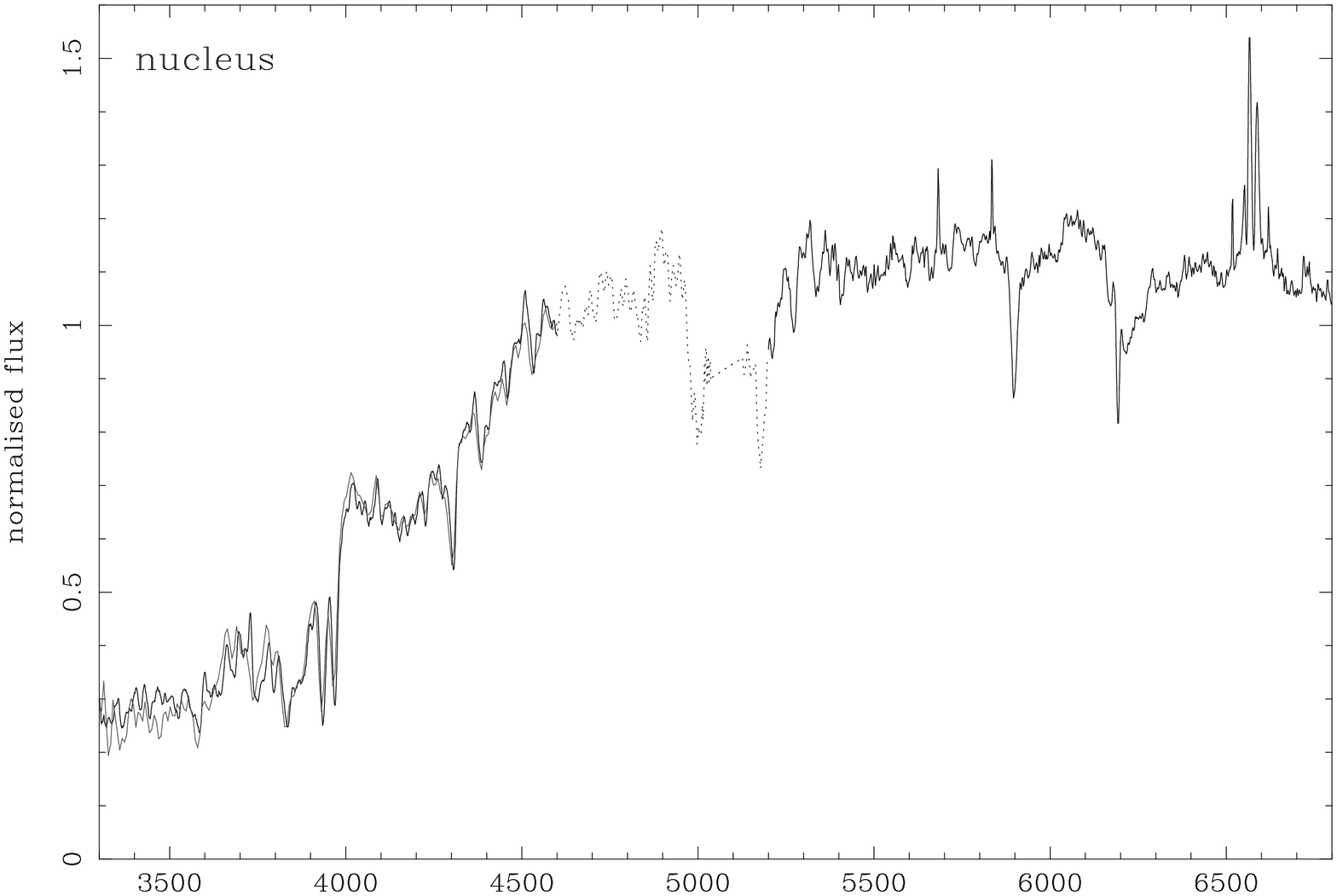}
\includegraphics[width=70mm,height=80mm,angle=270]{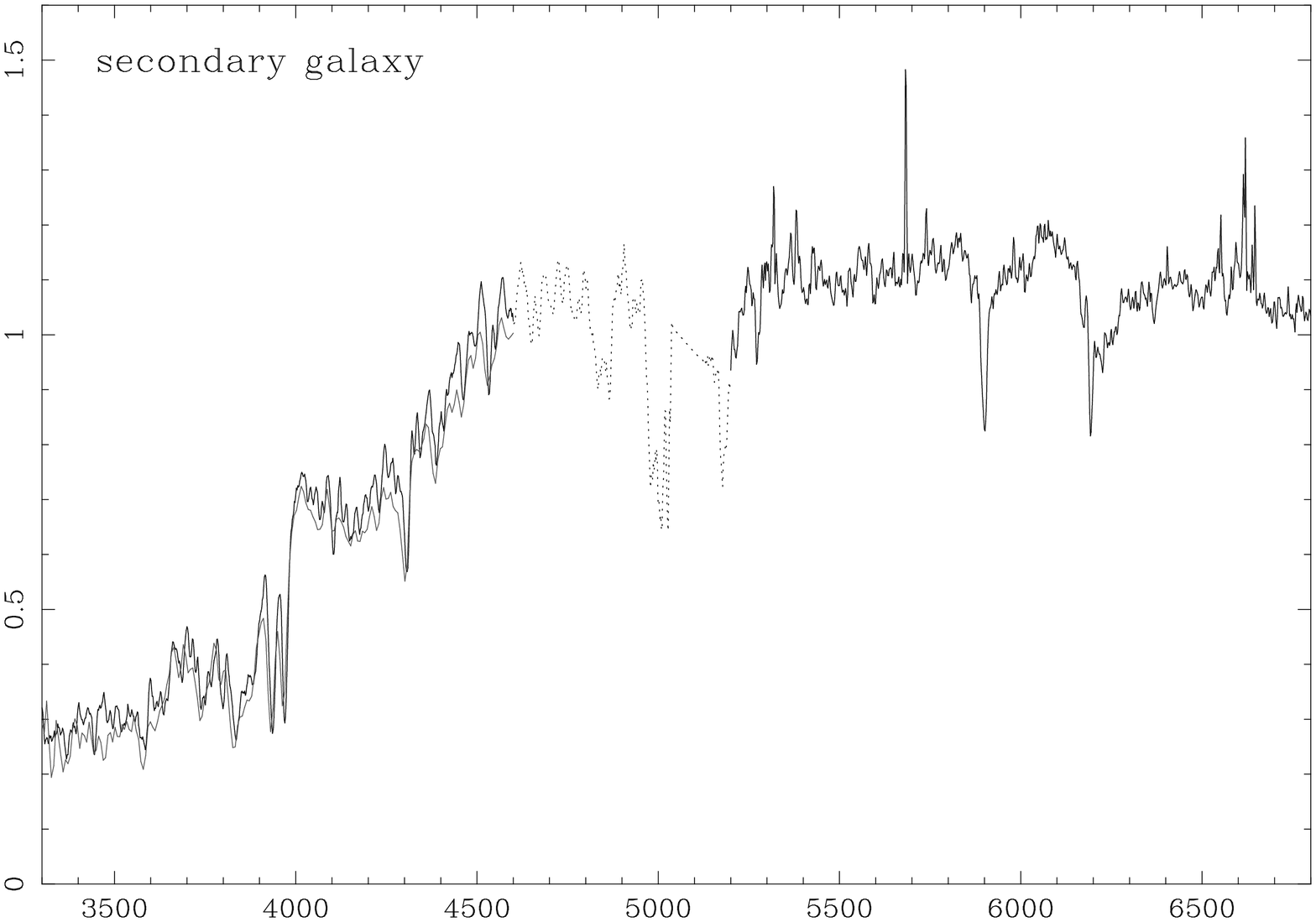}
\newline
\includegraphics[width=70mm,height=80mm,angle=270]{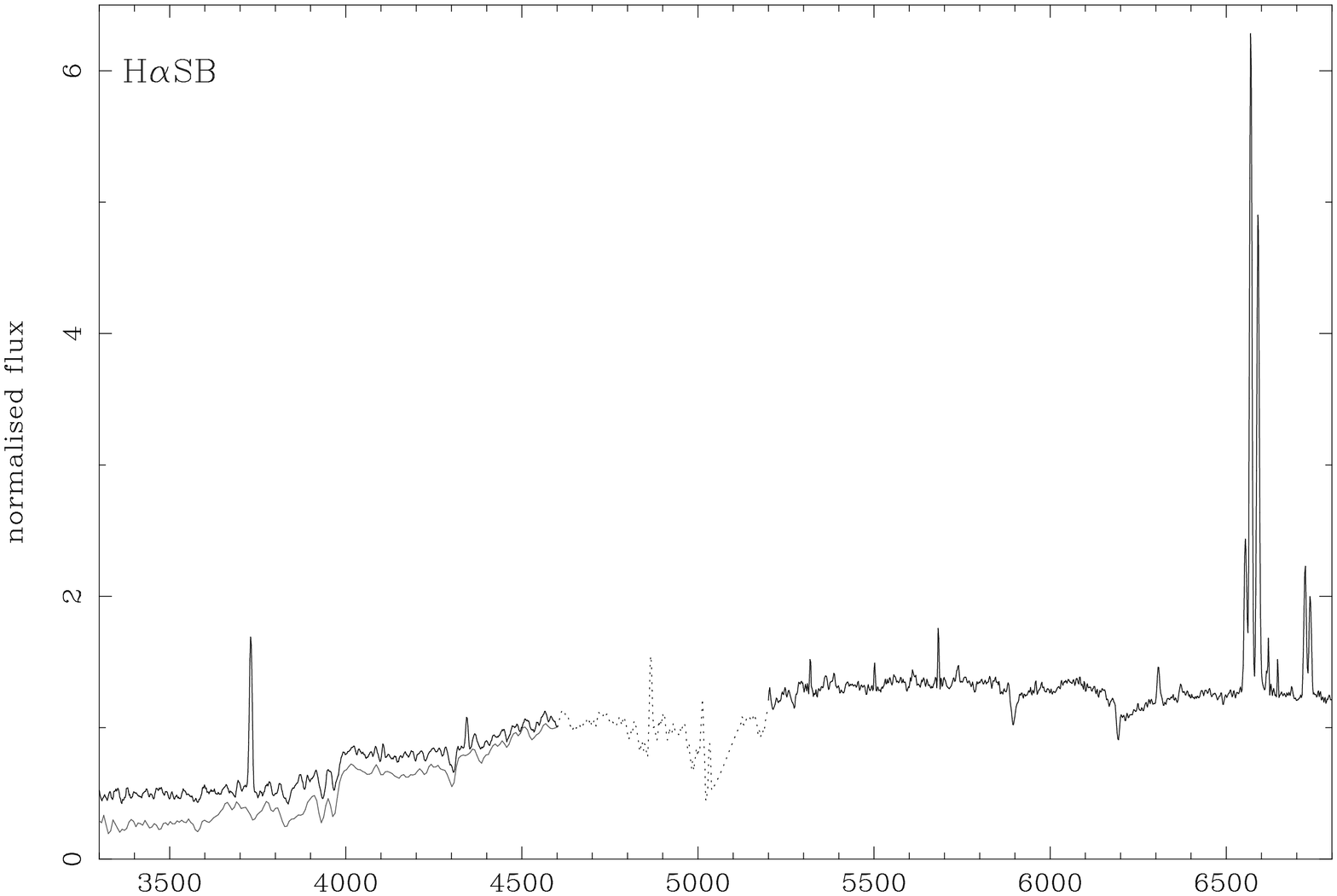}
\includegraphics[width=70mm,height=80mm,angle=270]{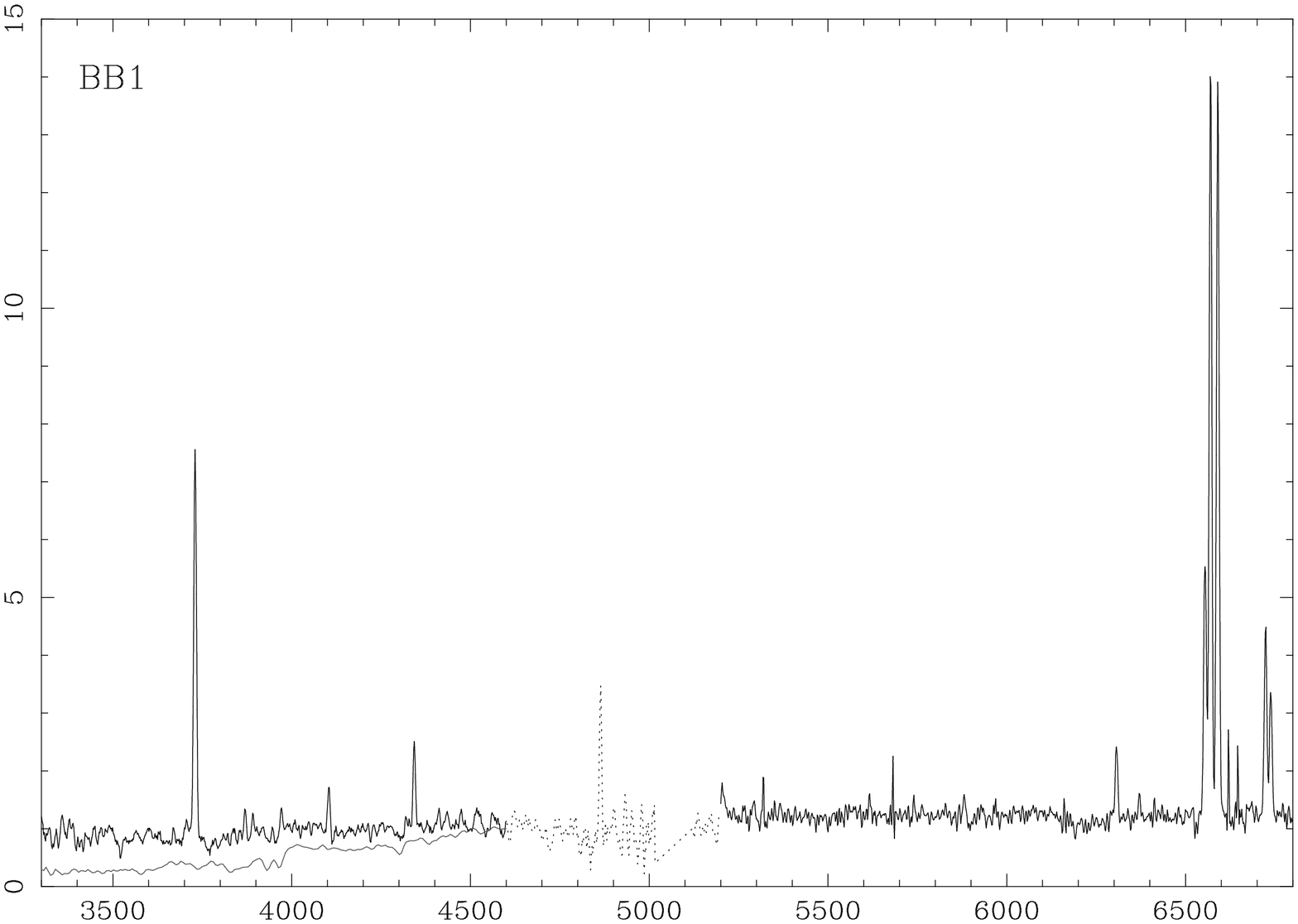}
\newline
\includegraphics[width=70mm,height=80mm,angle=270]{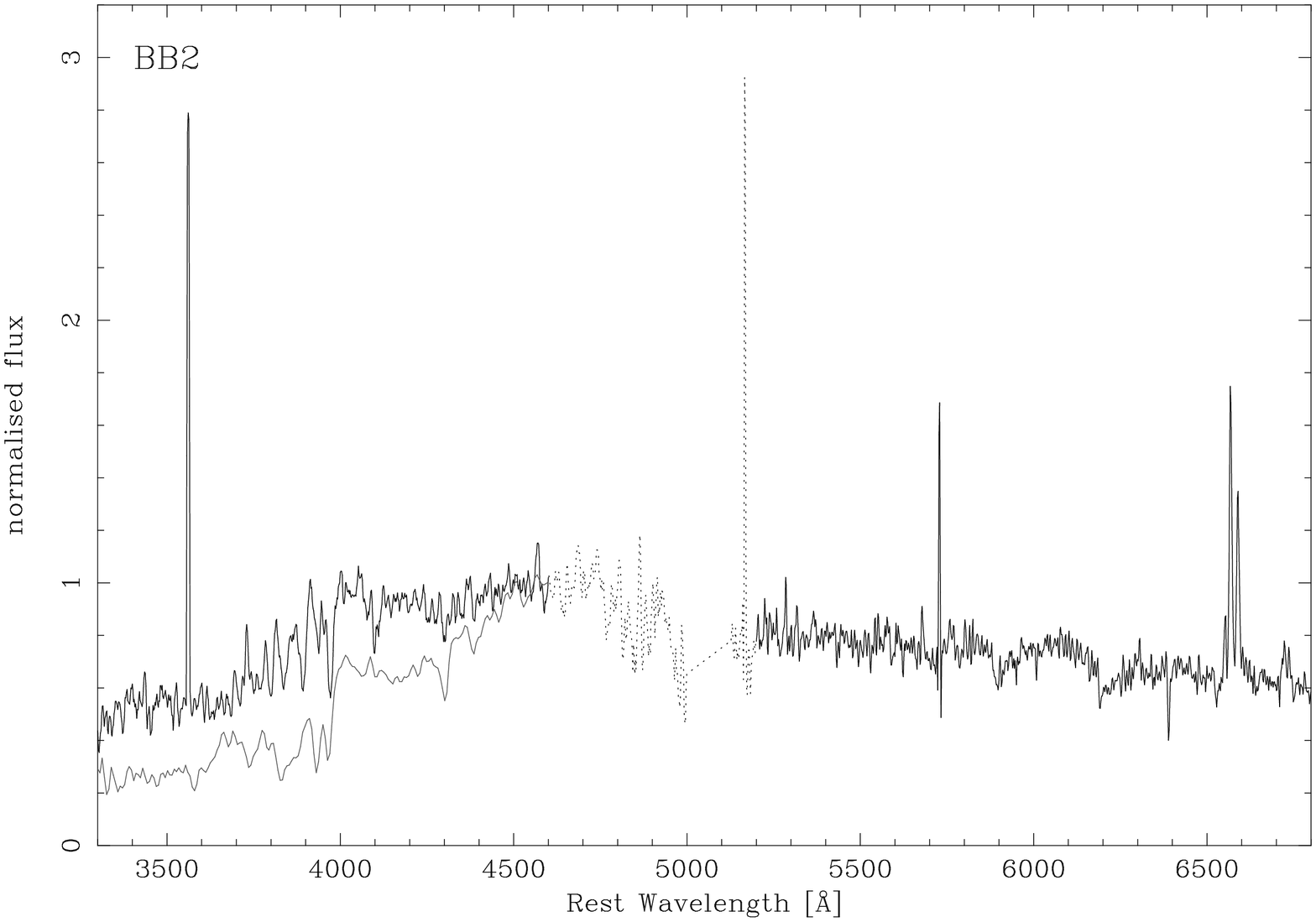}
\includegraphics[width=70mm,height=80mm,angle=270]{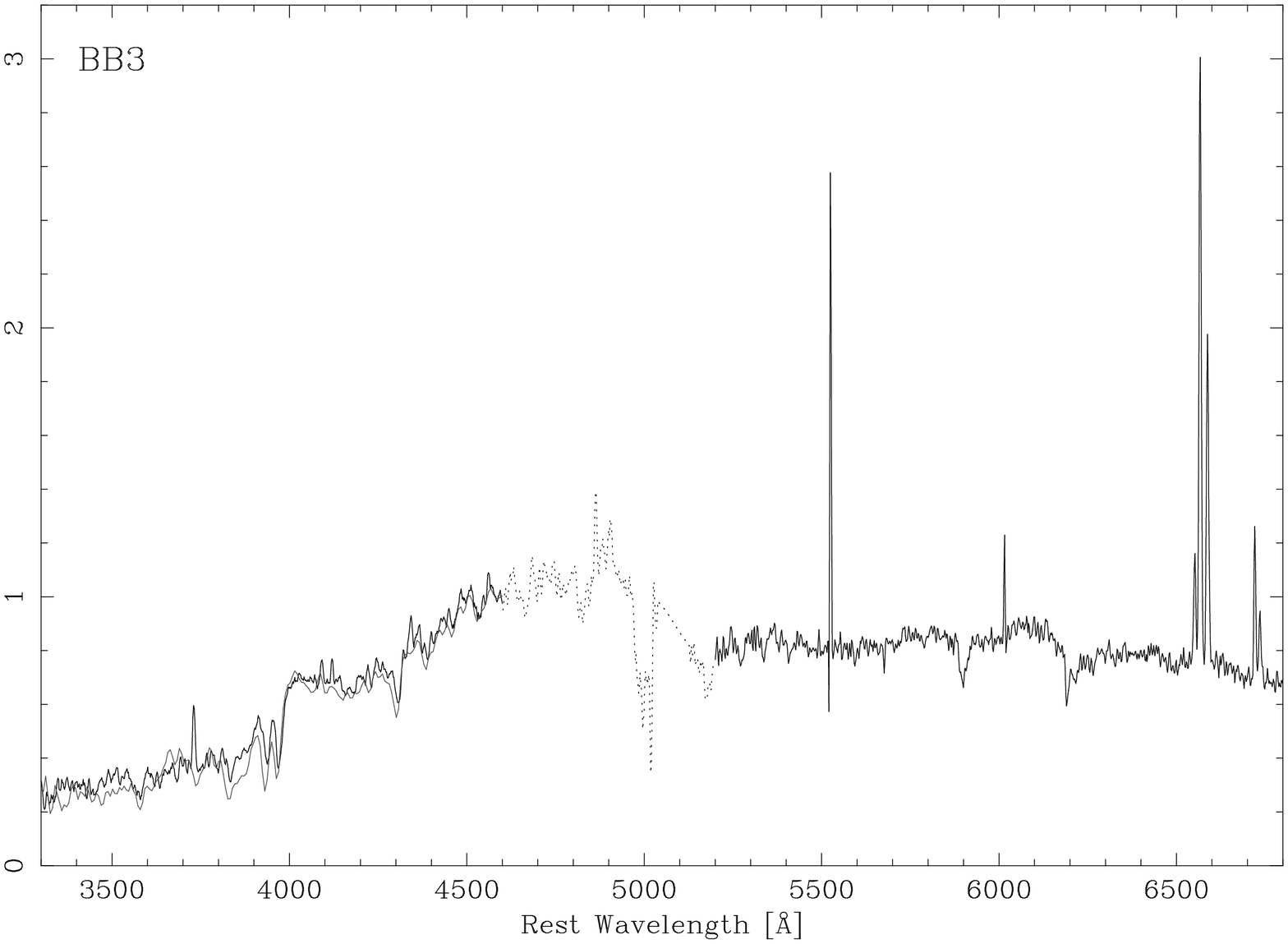}
\end{minipage}
\caption{Red and blue ISIS spectra of the main regions of interest around
RX~J0821; clockwise from the top left: the central cluster galaxy, the
secondary galaxy to the SE, BB1, BB3, BB2, and \ha SB. The spectra are
dotted in the region where the flux calibration is less accurate
because of the dichroic (between 4600-5200\thinspace\AA). The grey line
between up to 4600\thinspace\AA\ is a template central cluster galaxy spectrum
scaled to match the observed spectrum in the region around
4500\thinspace\AA. The spectra have {\sl not} been corrected for any
intrinsic reddening.}
\label{fig:unredspectra}
\end{figure*}

The secondary galaxy lying 4 \arc{} to the SE of the central cluster galaxy
is included in the PA305 observation, and has a spectrum typical of an
elliptical galaxy, lacking any significant line emission
(Figure~\ref{fig:unredspectra}). The continuum features show it to be
marginally redshifted with respect to the central cluster galaxy
($+77\pm 32$\kmps; see Sec. \ref{discussion}). PA305 also
covers a bright isolated blob of light seen in the HST image to lie
just inside the dark `bite' in the faint diffuse emission, 4.5 \arc{}
to the NW of the central galaxy (this is hereafter denoted as `BB1').
PA258 samples two much brighter clumps of light (hereafter `BB2' and
`BB3') that are representative of those forming the two arcing lines
crossing over to the North of the galaxy as seen in the HST image.
Finally, we also isolate a region (denoted `H$\alpha$SB' hereafter) in
PA305 that is located within the brightest part of the extended \ha\
emission nebula. Apart from its high H$\alpha$ luminosity, the
spectrum of this region is not very different to those of the other
blue regions. There is only a very weak counterpart to H$\alpha$SB in
the HST image. Figure~\ref{fig:unredspectra} shows the combined red
and blue arm spectrum of each of these regions, as well as those of the
central and secondary galaxies. The dashed line represents the blue
half of the spectrum of a template central cluster galaxy (see
Sec. \ref{continuum}) scaled to
match the spectrum in the region around 4500\thinspace\AA, illustrating the
excess of blue continuum light present in at least three of the blobs
(see also Fig. \ref{blobfits}).

\subsection{Emission lines}

The emission lines were fit in QDP (Tennant~1991) using composite
Gaussian models for the lines and a linear function to represent the
continuum. \oid, \niid, \siid, and the H$\alpha$ line were fit as a set of
7 Gaussians fixed to have the same redshift and velocity width as each
other (and the flux of [NII]$\lambda 6548$\thinspace\AA{} was fixed to
1/3 the value of [NII]$\lambda 6584$\thinspace\AA{}). Fitting the
bluer emission lines was more problematic as they are almost
completely extinguished, presumably by the reddening caused by large
amounts of intrinsic dust in parts of this system.
\hb\ was fit as an  individual Gaussian, with its velocity width fixed
to the same value as measured for the H$\alpha$ emission from that
spatial region. However, we had to allow for the possibility that
underlying stellar absorption had also contributed to this extinction,
especially if young stars might be present in significant quantity.

As a first approximation, we simply represented any stellar absorption
feature by including an underlying Gaussian absorption component into
the fit where a significant continuum level was present whose
properties (i.e. width) were matched from the Ca~H~\&~K lines. This
enabled us to attempt a conservative estimate for the
amount of intrinsic absorption to be corrected for in each spectrum.
The fit to \hb\ is further complicated, as due to the redshift of
RX~J0821, it lies in a region where the signal is beginning to be
affected by the dichroic used to split the light between the two arms
of ISIS. Where no Balmer lines other than H$\alpha$ could be detected,
a $2\sigma$ upper limit for H$\beta$ was derived.

H$\alpha$ is detectable in our slit spectra out to a distance of 9
\arc{} ($\sim24$\kpc) to the NW of the nucleus in slit PA305 with a
total luminosity of about $1.9\times 10^{41}$\ergps. Assuming
an average intrinsic reddening of {\sl E(B-V)}$=0.78$ (as determined from
the Balmer decrement; see Sec. \ref{extinction}) over the whole
slit, the corrected total H$\alpha$ slit luminosity
is about $1.26\pm 0.03\times 10^{42}$\ergps. Our slit encompasses
a maximum of app. 25\% of the total \ha\ flux in the TTF image, so we
estimate that the total \ha\ luminosity is at least as high as
$5\times10^{42}$\ergps. We note that the object's L(H$\alpha$)/L$_X$ ratio,
although not quite anomalous, is certainly very high as compared to the
other BCS objects. Of all the objects with comparable L$_X$, only one
(Zw 8193) has a similar H$\alpha$ luminosity.

The regions BB3 and H$\alpha$SB show particularly strong H$\alpha$
emission, whereas BB1 and BB2 are not conspicuous (Fig. \ref{lines},
central panels). However, before the correction for internal extinction
is applied (see next section), a distinct peak in H$\alpha$ emission
at the position of BB1 is apparent. Given the large error on the
derivation of {\sl E(B-V)} here, this peak in \ha\ could have been
lost (see measurements before de-reddening in Fig. \ref{lines}, centre
panels). BB2, however, seems to be intrinsically H$\alpha$-weak, which
agrees with the fact that other spectral properties differ between BB1
and BB2. The extremely high value for the \ha\ luminosity from BB3 is
due to the very high {\sl E(B-V)} at this position; this is the one
bin at this PA where the reddening determination seems to be fairly
secure.

Finally, there is a weak maximum in H$\alpha$ along slit PA305
approximately 7 \arc{} to the NW, which we termed the `Weak Blob' (wB)
as it is not apparent on the HST image. Its
H$\alpha$ luminosity and ionization properties (see Sec
\ref{ionization}) suggest
that it might be another candidate blue blob.

\subsubsection{Intrinsic extinction}
\label{extinction}
The previous spectrum of RX~J0821 (Crawford \etal 1995) had already
suggested a very high amount of intrinsic reddening to be present, and
our WHT data immediately confirm the
importance of dust in this object. We compared our measured Balmer
decrements to values predicted by Case B recombination, to derive a
value for {\sl E(B-V)} for each of our spatial bins, as the
dust is most likely not distributed in a uniform screen. An additional
independent determination of {\sl E(B-V)} was performed using the
H$\alpha$/H$\gamma$ ratio in those spectral bins where a stronger
Balmer series was visible. 

\begin{figure*}
\begin{minipage}{175mm}
\includegraphics[width=70mm,height=85mm,angle=270]{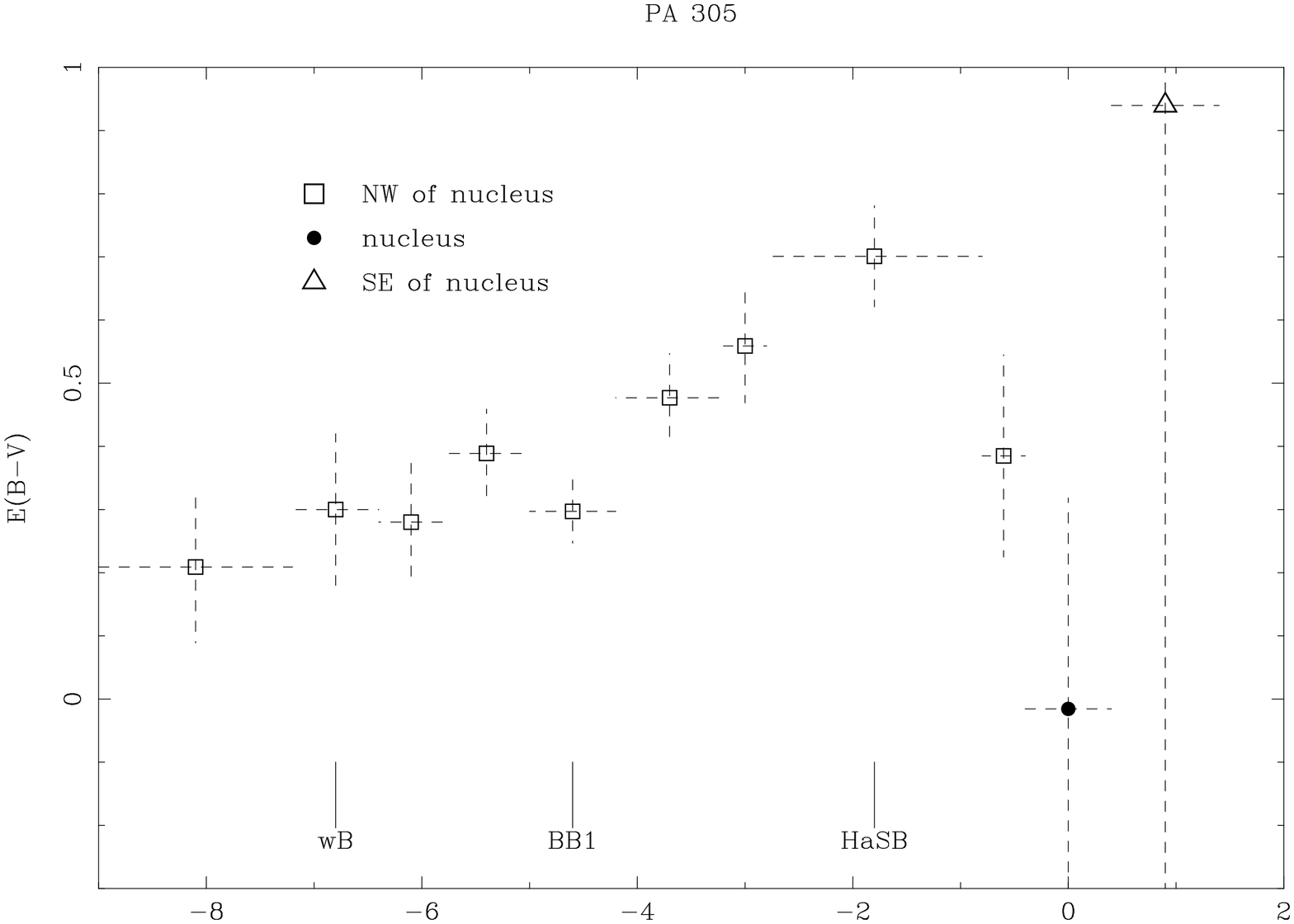}
\includegraphics[width=70mm,height=85mm,angle=270]{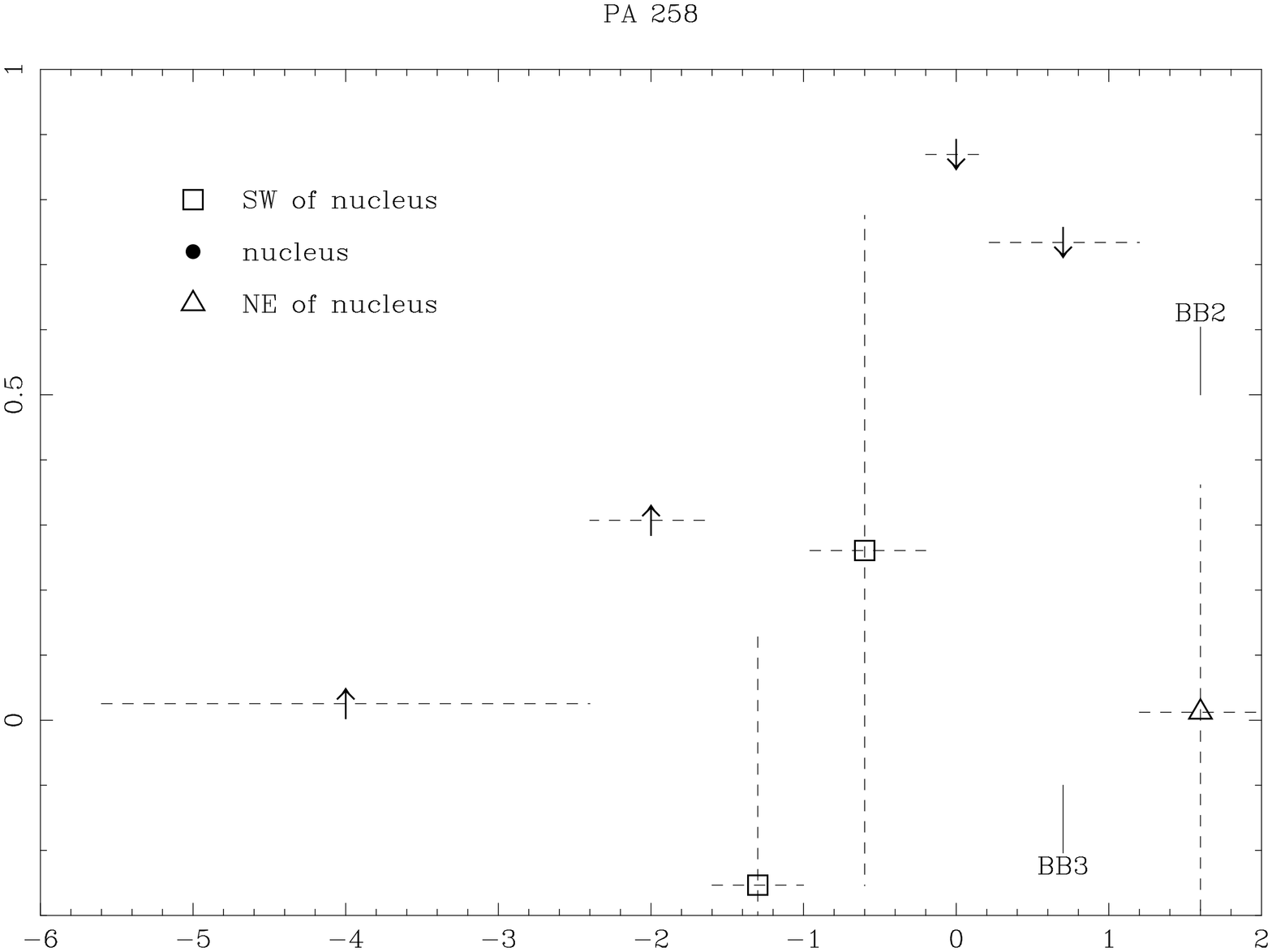}
\newline
\includegraphics[width=70mm,height=85mm,angle=270]{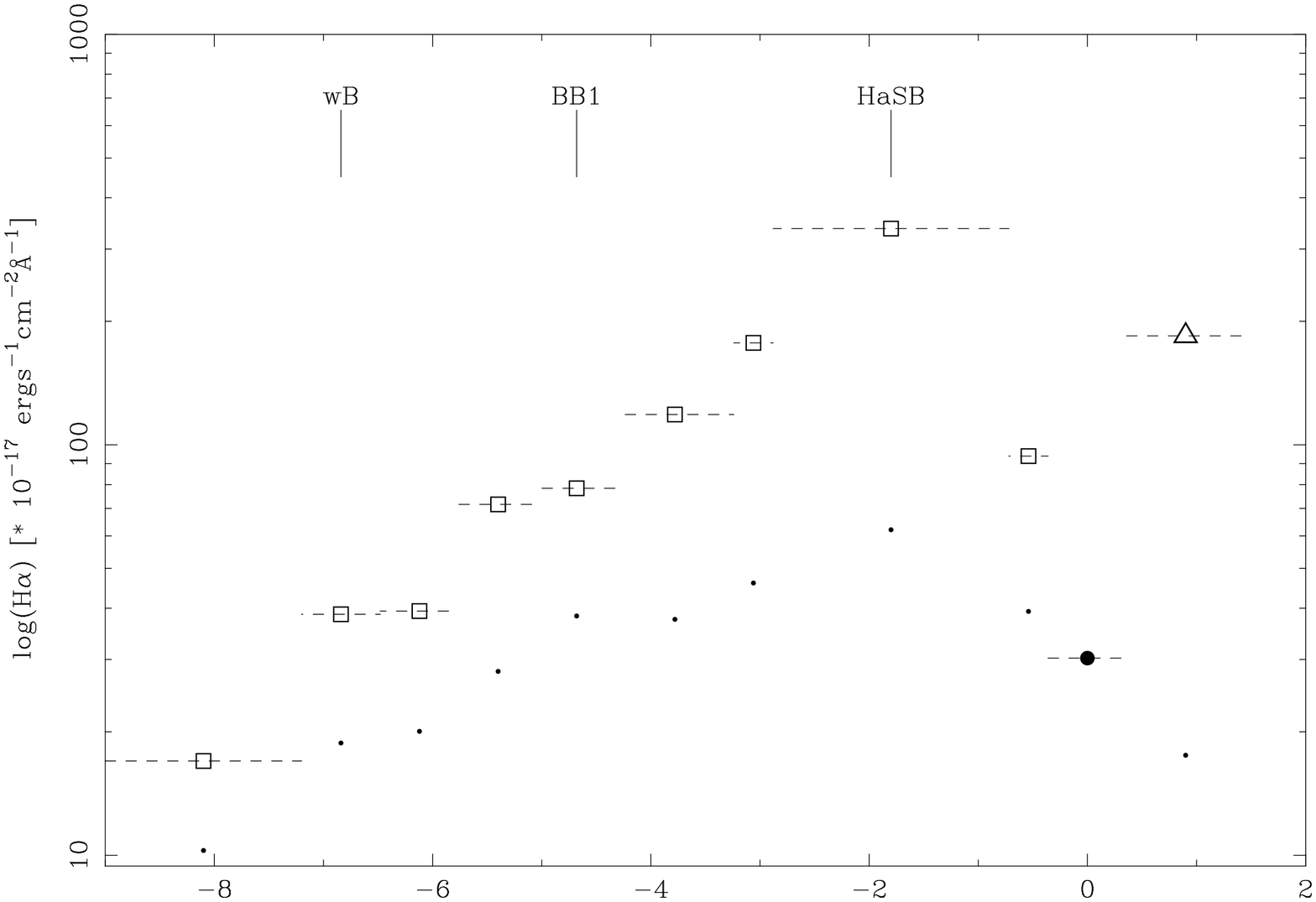}
\includegraphics[width=70mm,height=85mm,angle=270]{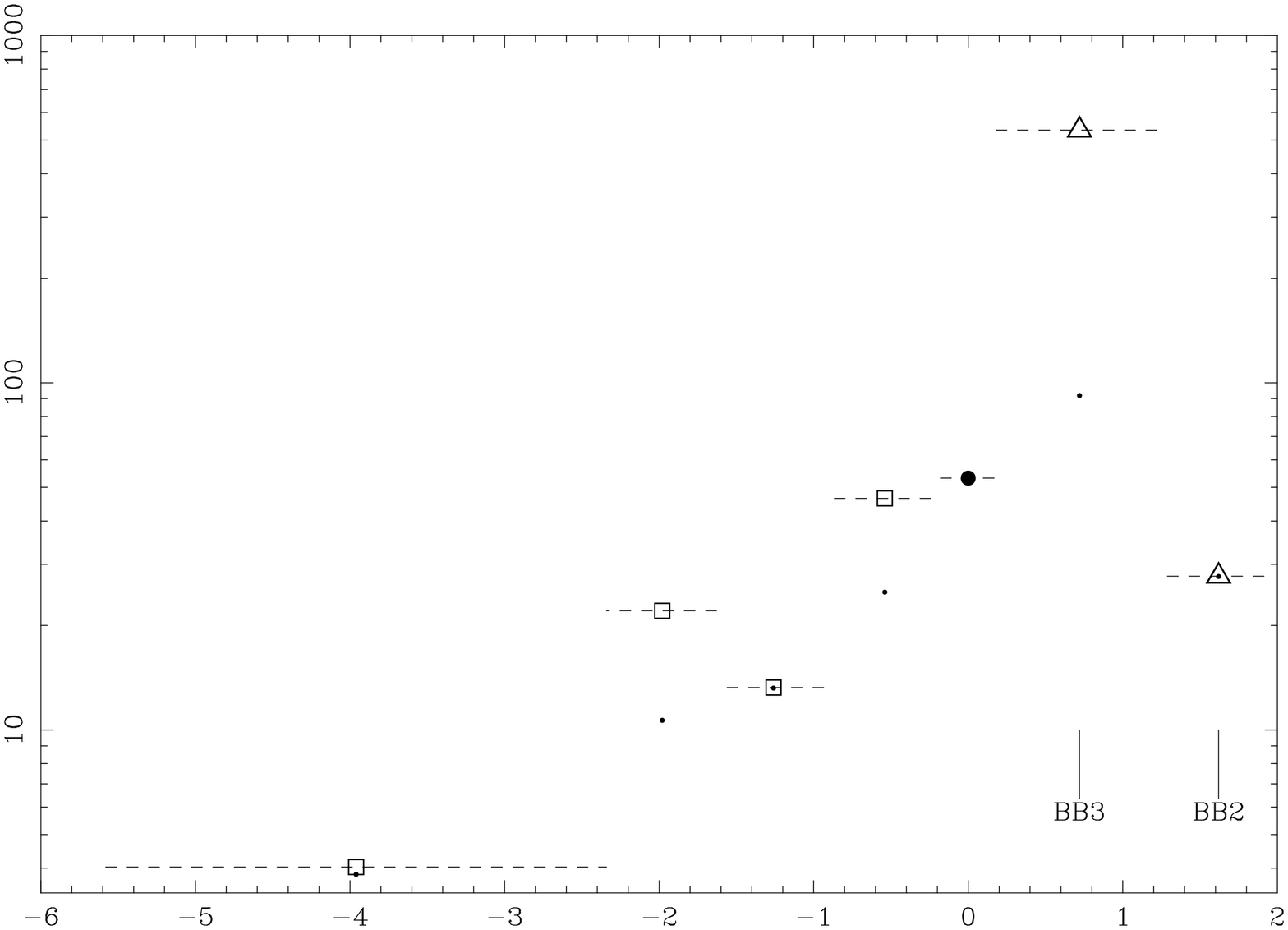}
\newline
\includegraphics[width=70mm,height=85mm,angle=270]{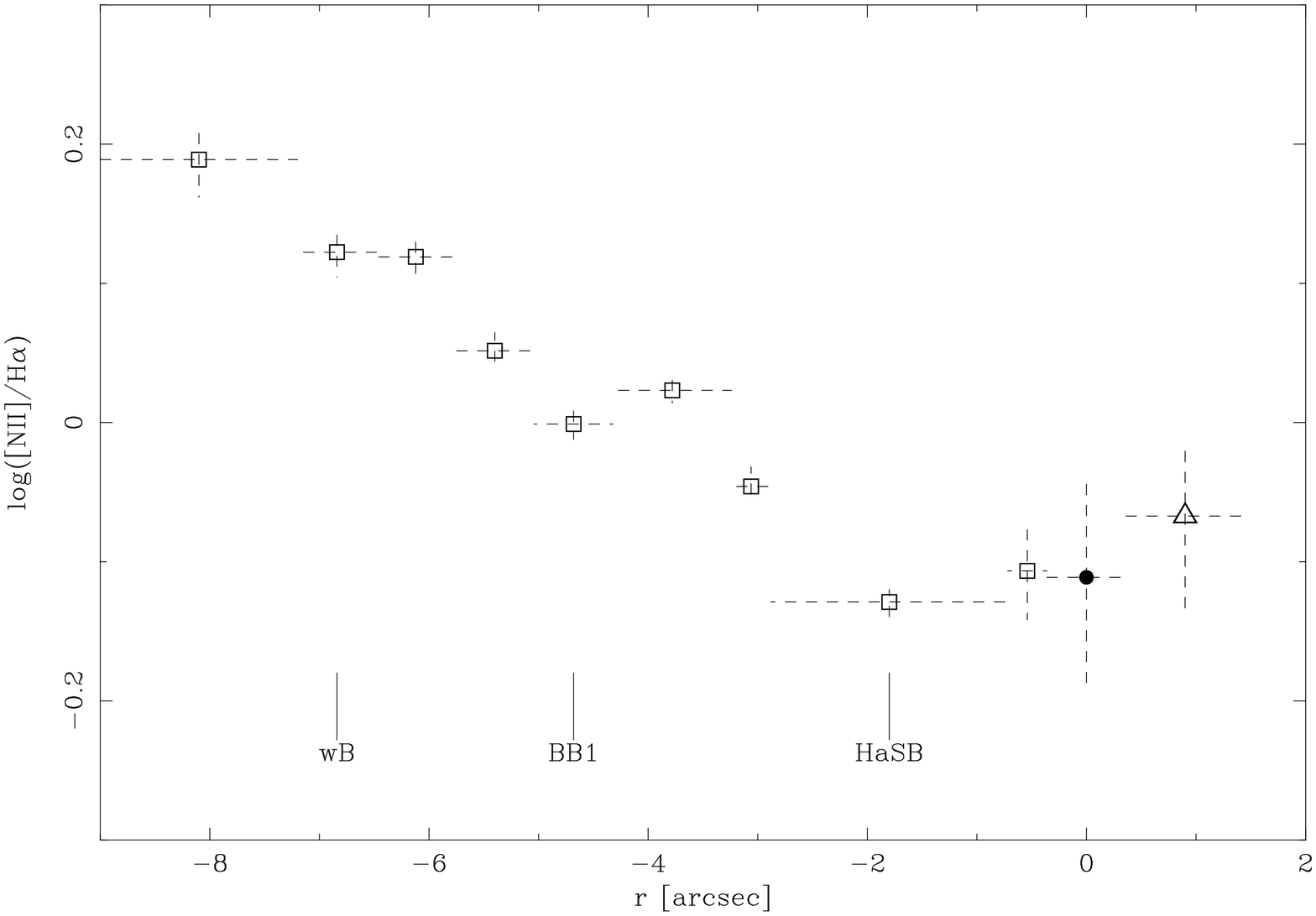}
\includegraphics[width=70mm,height=85mm,angle=270]{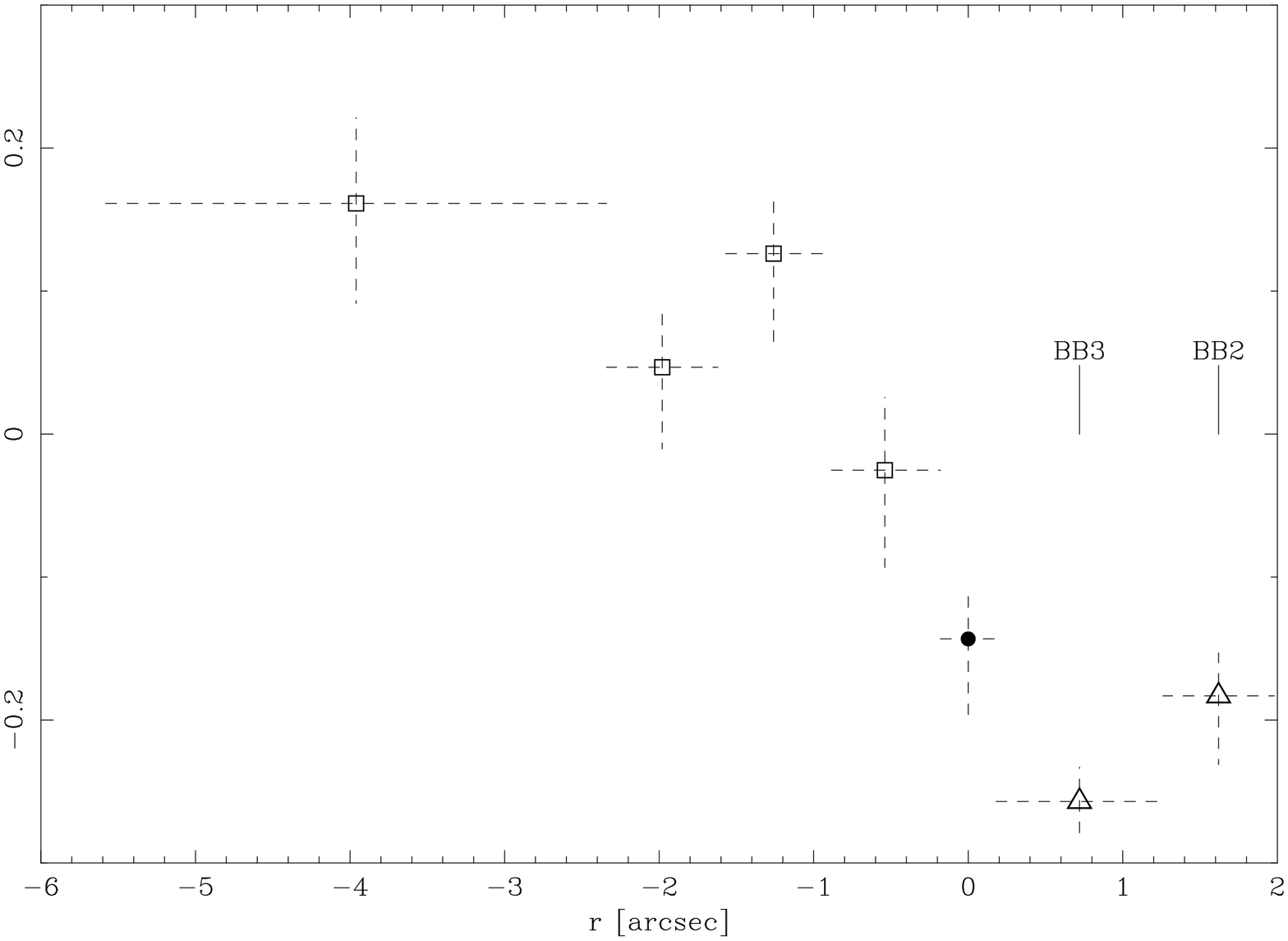}
\end{minipage}
\caption{ Plots of the intrinsic {\sl E(B-V)} measured from the ratio of
\ha/\hb\ (top), the surface brightness
in \ha\ where the dots denote values before de-reddening (middle),
and the line intensity ratio
[NII]/\ha\ (bottom) as a function of radius from the central cluster
galaxy (the centre of which is marked by a solid circle in all plots).
The spectral properties are shown for slit PA305 (left) and PA258
(right). Arrows indicate upper (cf text) and lower limits, and the
approximate positions of
the regions referred to in the text are marked. Errors on the
H$\alpha$ measurements are not represented here as they are so small
that for most bins they would be within the symbol.
}
\label{lines}
\end{figure*}

Not surprisingly, given the patchy structure visible in the HST image,
the amount of reddening varies considerably along the slit. In the
upper panels of Fig.~\ref{lines}, the inferred {\sl E(B-V)} is mapped
as a function of radial distance from the central galaxy at each slit
position angle. These values of {\sl E(B-V)} are either weighted
means of all available measurements, or, where only H$\alpha$ was
visible, lower limits. The errors shown on {\sl E(B-V)} are propagated
from the $\Delta\chi^2=2.71$ errors ($=90\%$ confidence interval) to
the fits of the Balmer emission
line fluxes in QDP. In PA305 reddening is unimportant around the
nuclear region, with a steep rise to the NW, peaking around the
position of the \ha SB, consistent with the edge of the dark `bite'
seen in the HST image. The {\sl E(B-V)} then decreases continually
further to the NW, with the blue blob BB1 perhaps located at a local
minimum (though this dip is probably not significant).

In PA258 a very high value for {\sl E(B-V)} in the nucleus suggests
that either this slit position is not centred on the same bit of the
main galaxy as the PA305, or that there is a comparative loss of blue
light due to atmospheric refraction (and thus the amount of \hb\ is
underestimated relative to \ha\ and the reddening inferred is
overestimated). The parallactic angle at the time of the observations
was about 309\deg, so we expect that only PA258 could possibly be
affected by this. There is evidence that reddening is insignificant in
the nucleus (see Sec \ref{modelresults}), so we assume the {\sl E(B-V)}
derived for this region in PA258 to be an upper limit.
Similarly, although the data suggest that BB3 seems to be associated
with a region of
intrinsically very high reddening, the large errors on this derivation
and the possibility of a slight comparative loss of blue light at
PA258 suggest we should employ caution and use this value of {\sl
E(B-V)} only as an upper limit. 

The average value of {\sl E(B-V)} for the
whole object as determined from a sum of all the available hydrogen
Balmer light was found to be $0.78^{+0.09}_{-0.11}$. This value agrees
within the error bars with that given in Crawford \etal (1999; {\sl
E(B-V)}$=1.16^{+0.33}_{-0.58}$).

\subsubsection{Ionization}
\label{ionization}
\begin{figure}
\centering
\includegraphics[width=0.3\textwidth,angle=270]{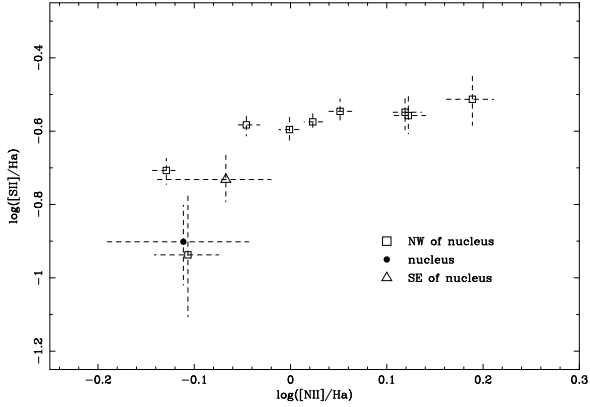}
\caption{Log([SII]/H$\alpha$) vs Log([NII]/H$\alpha$) for PA
305\deg. Note the strong, almost monotonic variation of the ratios.}
\label{s2vsn2}
\end{figure}

Whilst the high levels of intrinsic dust extinction strongly limit
our ability to measure the properties of the blue emission lines (such
as [OII], \hb+[OIII]), the strength of the red emission lines means
that these lines can be fit with some precision. We fit and measure
the red diagnostic line intensity ratios [NII]$\lambda$6584/H$\alpha$,
[SII]$\lambda$6717/H$\alpha$ and [OI]$\lambda$6300/H$\alpha$, which
are relatively unaffected by the local variations in reddening. Marked
changes in ionization across RX~J0821 are obvious (Fig. \ref{s2vsn2}). 

At the location of the blobs, the ionizational behaviour deviates from
the general trend, in that they show more extreme values in
ionization-sensitive ratios than their immediate environments
(Fig.~\ref{lines}, lower panels). All blue blobs (including the Weak
Blob wB; but
with the possible exception of BB2) are characterised by a clear local
minimum of the [NII]/H$\alpha$ (and also the [SII]/H$\alpha$ and
[OI]/H$\alpha$ ratios).

\begin{figure}
\centering
\includegraphics[width=0.3\textwidth,angle=270]{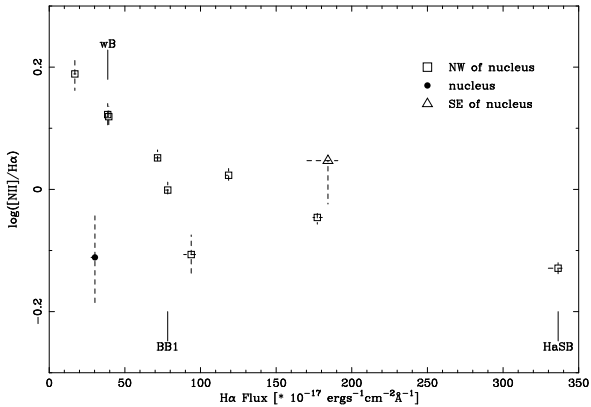}
\caption{Log([NII]/H$\alpha$) vs H$\alpha$ for PA 305\deg. The blobs
are indicated.}
\label{n2vsha}
\end{figure}

There is also a clear correlation between the H$\alpha$ flux and
[NII]/H$\alpha$, such that stronger H$\alpha$ is associated with a
lower value of [NII]/H$\alpha$; \ie regions where photoionization is
more important (lower [NII]/H$\alpha$) are stronger H$\alpha$ emitters
(Fig.~\ref{n2vsha}). PA258 confirms the trend found in PA305, but is
not included in Fig.~\ref{n2vsha} due to its lower S/N. The exception
to this correlation (for both slit position angles) is the region
around the galaxy nucleus. Interestingly, this correlation matches the
general trend found from object to object in the BCS (Crawford \etal 1999,
fig. 7). 

\subsubsection{Kinematics}
\label{kinematics}
\begin{figure*}
\begin{minipage}{175mm}
\includegraphics[width=70mm,height=85mm,angle=270]{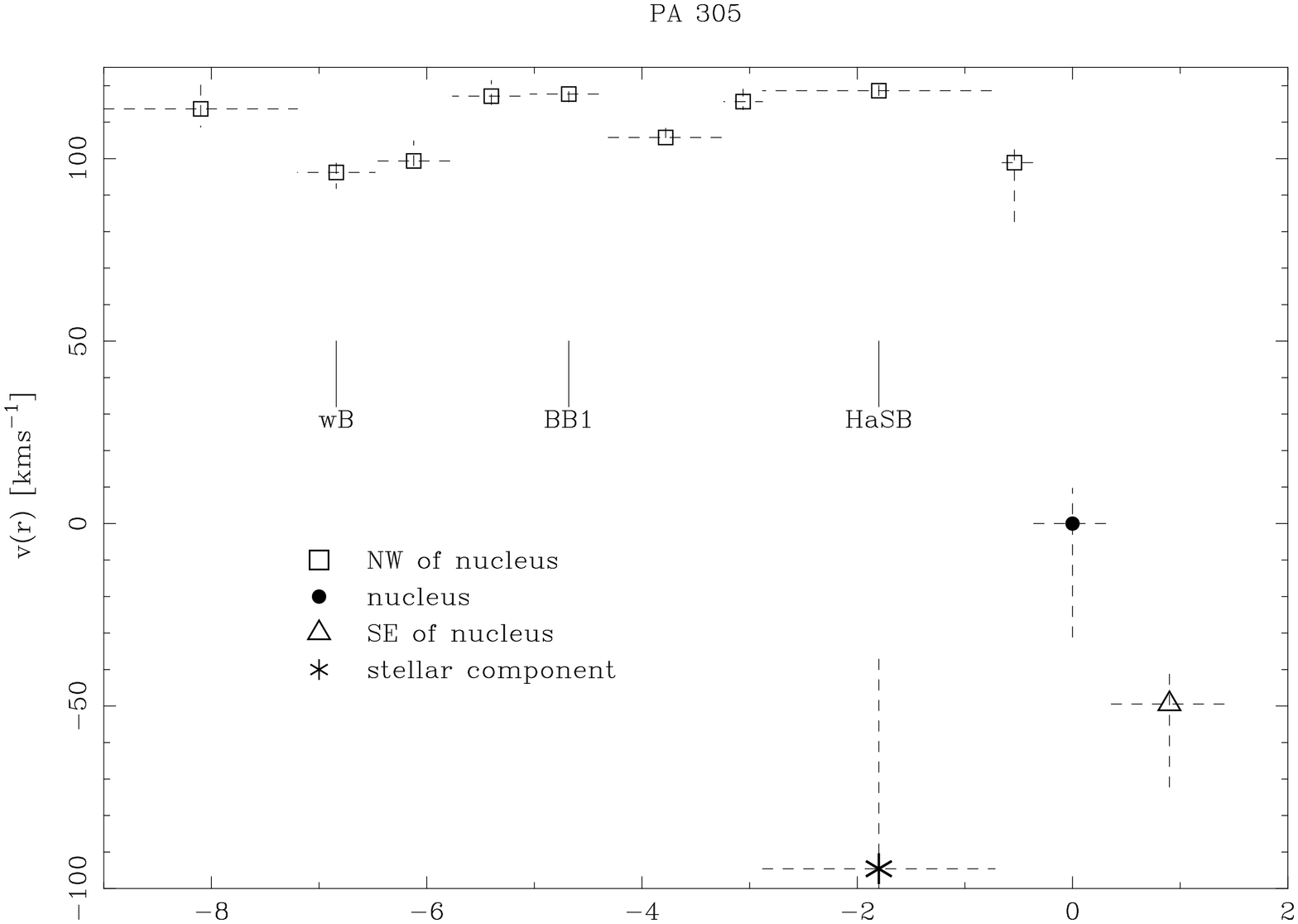}
\includegraphics[width=70mm,height=85mm,angle=270]{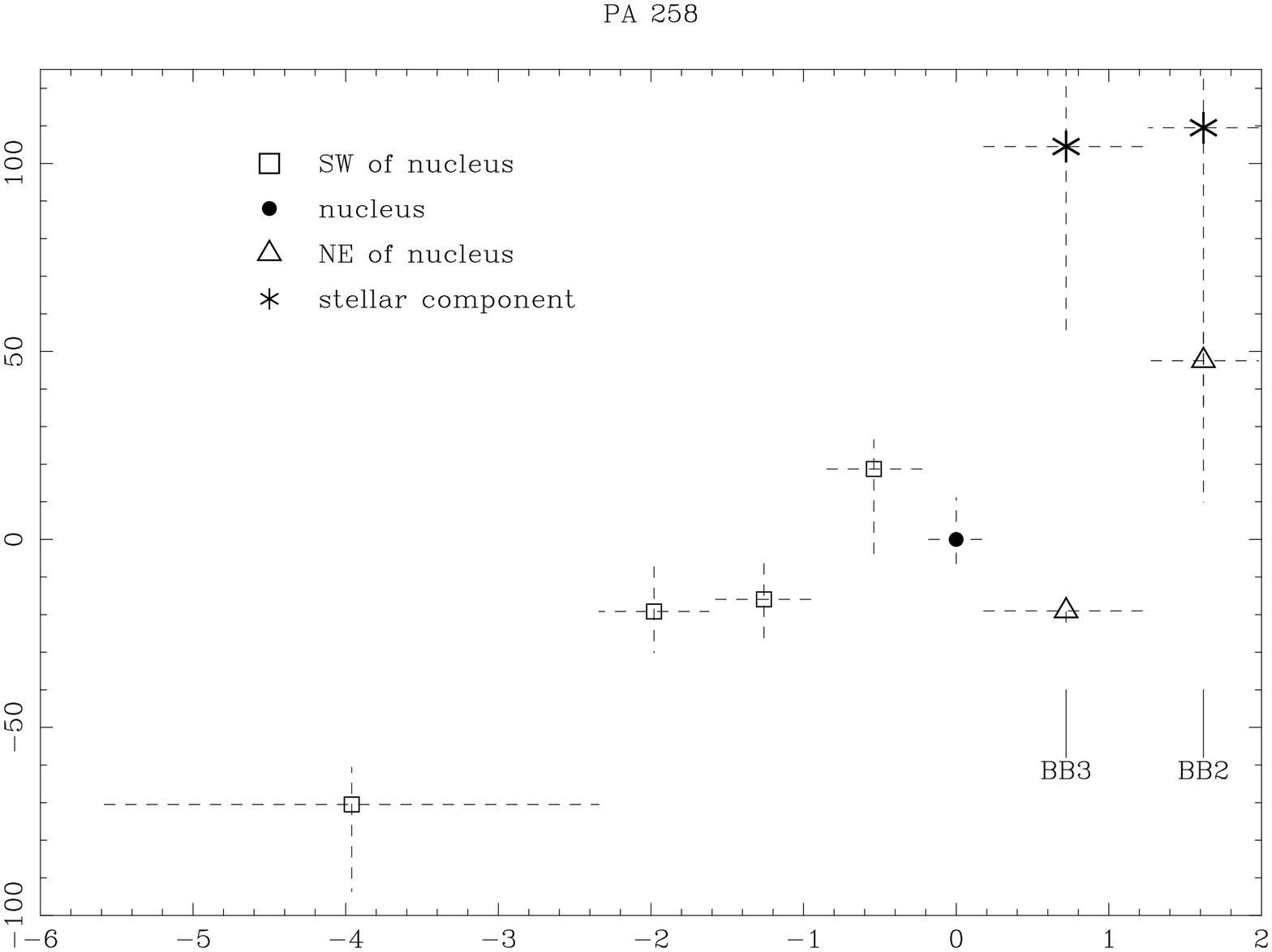}
\newline
\includegraphics[width=70mm,height=85mm,angle=270]{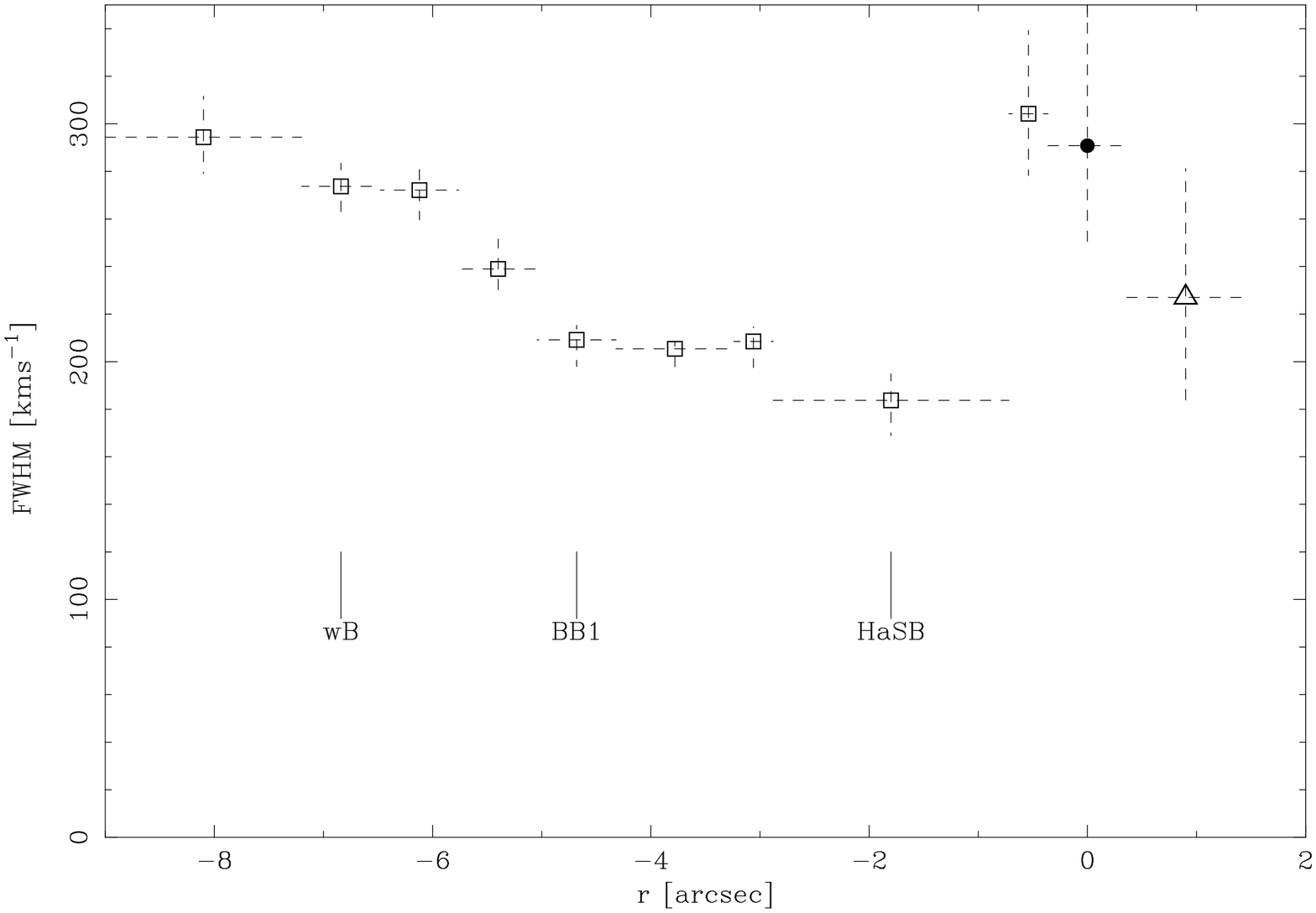}
\includegraphics[width=70mm,height=85mm,angle=270]{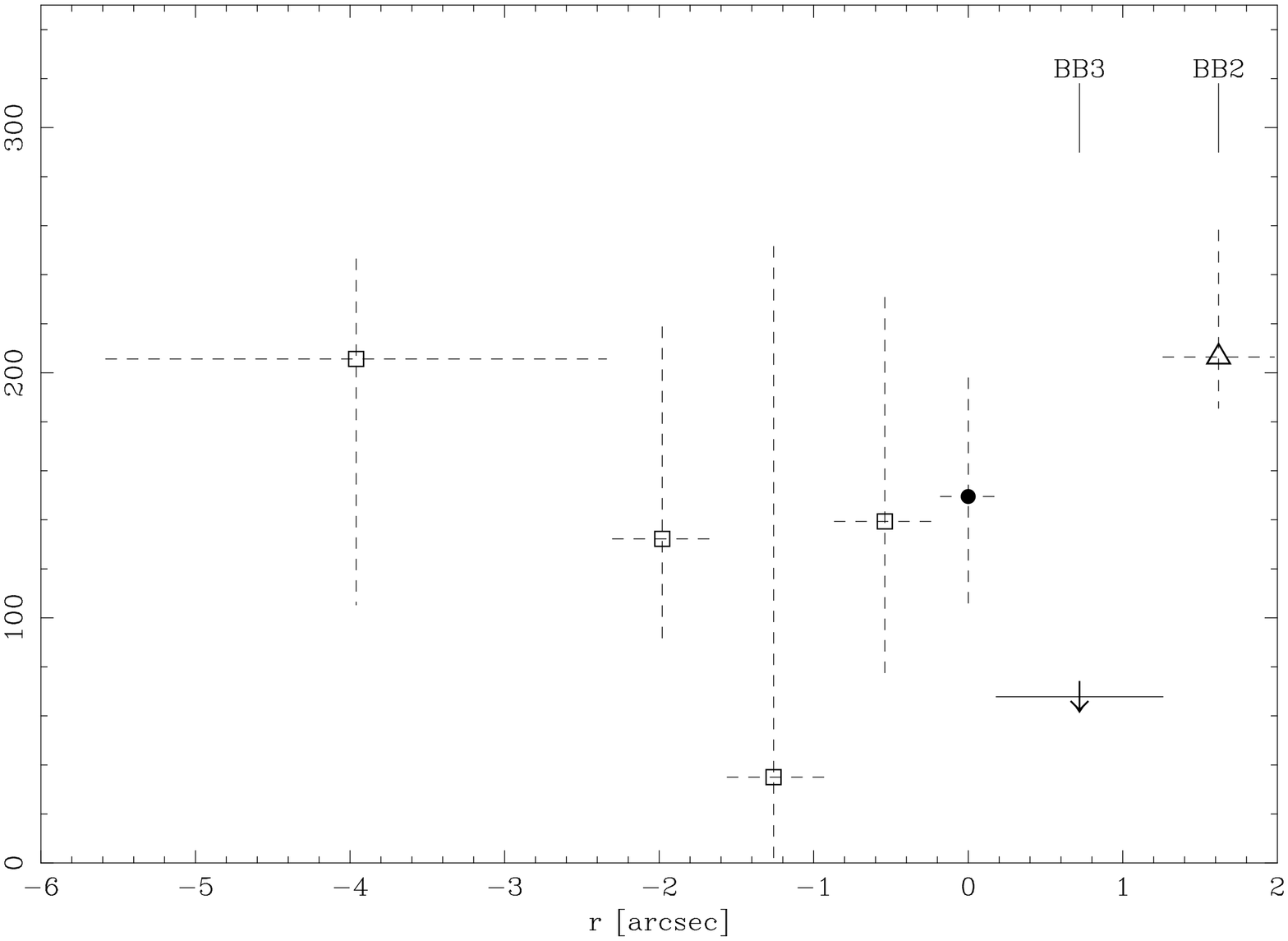}
\end{minipage}
\caption{Kinematic properties of the emission-line gas. Arrows
indicate upper limits, and the
approximate positions of the blobs are denoted. Asterisks denote values of
$v_{rad}$ derived from the Ca~H~\&~K absorption lines measured relative
to the absorption in the nucleus, representing the stellar component
of the blobs.}
\label{kin}
\end{figure*}

The QDP fits to the emission lines enable us to measure any velocity shifts
of the emission line nebula, as well as variations in line width
across the system. The upper panels in Fig.~\ref{kin} show the radial
velocity $v_{rad}$ (relative to the nuclear bin) of the
\ha\ as a function of radius for both slit position angles.
The $v_{rad}$ values measured from PA305 show a sharp initial increase
in $v_{rad}$ from SE to NW of around $150$\kmps over a distance of
only 2 \arc{} (5.5\kpc), which then levels off to small variation
around +100\kmps. To this NW section at least two kinematically
distinct regions can be distinguished: one local minimum of $v_{rad}$ at about
$r=-4$ \arc{}, another one at around $r=-6.5$ \arc{}. These locations
coincide well with positions {\sl in between} H$\alpha$SB and BB1, and
BB1 and wB, respectively, and are associated with local maxima of
[NII]/H$\alpha$. As a corollary, these blobs themselves are associated
with the higher values of $v_{rad}$. The worse S/N
in the PA258 slit means that it is difficult to draw any firm
conclusions about the relative kinematics of BB3 and BB2: BB3 may be
also associated with a local extremum of $v_{rad}$, but the lack of
further line emission ENE along the slit makes this difficult to assess.

The asterisks in Fig.~\ref{kin} mark the values of
$v_{rad}$ derived from the Ca~H~\&~K absorption lines to check for any
kinematic difference between the stellar and gaseous components. The
stellar component associated with the H$\alpha$SB is {\sl blueshifted}
relative to the nucleus by about the same amount the H$\alpha$
emission is {\sl redshifted}. Due to the
weakness of the Ca~H ~\&~K lines in BB1, the error bars here are very
large and the
result has not been included in the plot. However, the result of
$v_{rad}\sim -40$\kmps\ is at least consistent with a similarly
strong velocity offset as \ha SB, confirming some degree of
independence between the gas and stars NW of the galaxy. 
A similar but less significant kinematic
separation is observed in BB3, and there might also be an even smaller
velocity offset in BB2, although the
velocities here are still consistent with each other within the error
bars. Measurements of $v_{rad}$ using [OII]$\lambda 3727$\thinspace\AA\
rather than \ha\ confirm this qualitative picture. We note that this
can be interpreted both in terms of a spatial, as well as purely
kinematical separation.

\begin{figure}
\centering
\includegraphics[width=0.3\textwidth,angle=270]{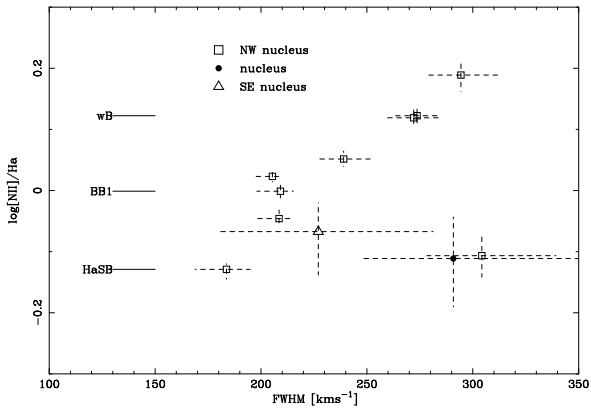}
\caption{log([NII]/H$\alpha$) vs FWHM for PA 305\deg.}
\label{n2vss}
\end{figure}

The lower panels of Fig.~\ref{kin} show the linewidth measured from \ha\
(corrected for instrumental resolution) as a function of radius for
both position angles. The discrepancy of the nuclear FWHM for the two
different slit orientations could be another indication that the slit
was not centred on exactly the same position during the two
observations (see Sec. \ref{extinction}). The blobs seem to be
associated with regions of
systematically smaller linewidth, although again the lower S/N in
PA258 renders it more difficult to judge the kinematic behaviour
around BB2 and BB3. Even so, there is a very low upper limit at BB3
that could indicate a real drop, and it appears that again BB2 is
different from the other blobs, having a larger FWHM. The nucleus is
generally associated with high values of FWHM. There is also a strong
correlation between the FWHM and [NII]/H$\alpha$ to the NW of the
galaxy (Fig. \ref{n2vss}), in that the regions showing higher
[NII]/H$\alpha$ have broader lines. As in Fig.
\ref{n2vsha} the nucleus and its immediate environment do not conform
to this trend. 

\subsection{The continuum}
\label{continuum}

The nuclear spectra of the main and secondary galaxies extracted from
PA305 are well fit by an average spectrum formed from the
spectra of 24 central cluster galaxies without line emission or excess
blue light (from Crawford \etal 1999; Figure~\ref{fig:unredspectra}).
All of the four other regions, however, are characterised by their
blue continuum in comparison to either the main galaxy nucleus or the
template central cluster galaxy spectrum (and hence their nomenclature
`blue blobs'). 
The blue excess is somewhat less noticable in BB3, with a spectrum that is
close in shape
to that of the nucleus of the central galaxy.
Note however that due to being so close to the bright galaxy
core ($r\approx 0.7$\arc{}), BB3 clearly is not fully resolved, and it is
very likely that
its spectrum is diluted by the spectrum of the nucleus itself (and
vice versa), and the
blue excess here could be attenuated. Further support for this is
provided in Sec \ref{modelresults}.

Closer inspection of the spectrum of BB2 (Fig.~\ref{blobfits})
shows that it has
a clear Balmer {\sl absorption} sequence (extending beyond H$\zeta$)
immediately suggesting a strong, relatively undiluted population of
A/F stars. 

This blue excess is apparent in the spectra as observed
(Fig.~\ref{fig:unredspectra}) but becomes more so after correction
for intrinsic reddening is made
(Fig.~\ref{blobfits}). The values of {\sl E(B-V)} adopted for H$\alpha$SB,
BB1, BB3, and BB2 are 0.70, 0.30, 0.00, and 0.01 respectively. Due to the
low S/N ratio in PA258, it was difficult to derive a sensible value
for {\sl E(B-V)} for BB3. Hence we use an assumption of {\sl E(B-V)}$=0$
rather than 0.73 (the value suggested by the Balmer decrement) as a
conservative minimum estimate, meaning that the real blue excess might be
significantly higher. 

In order to estimate the dominant stellar populations responsible for
the excess light, we carried out an empirical stellar spectral
synthesis, similar to that described in Allen (1995) or Crawford \etal
(1999).

\subsubsection{The tools}
\label{tools}

We fit the spectra from the different spatial regions of RX~J0821
by combinations of main sequence dwarf and giant stars, as
well as the featureless central cluster galaxy template. We aimed to
reproduce both certain
spectral features and the general spectral slope. We made use
of the `\tt{}specfit\rm{}' package in IRAF (Kriss~1994). This software
allows one to fit spectra with a range of pre-defined models as well as
user-defined ones, that can be fed into \tt{}specfit\rm{} in the form
of an ASCII table. The algorithm fits the user spectra to the input
spectrum with the normalization and -- if required -- the redshift as
free parameters. 
As it is principally the blue light that we wish to fit, and which
contains the  most strongly discriminating stellar spectral
features, we fit models to the blue-arm ISIS spectra only, over a
wavelength range of about $3300-4600$\thinspace\AA{}. 

\subsubsection{The models}
\label{models}

The template spectra used in `\tt{}specfit\rm{}' were taken from
Pickles' extensive stellar spectral flux library (Pickles 1998), and
the template central cluster galaxy spectrum constructed from
24 featureless central cluster galaxies (Crawford \etal 1999). We have
confidence in the validity of this template, as it only requires a
simple normalization to be a good fit to the nuclear spectrum of
RX~J0821 -- where our Balmer line ratios indicate no
intrinsic reddening to be present. The fit is very good,
confirming that the centre of this object is consistent with a normal,
elliptical galaxy, and that the determined reddening of {\sl E(B-V)}
$\sim 0$ is a realistic value. The red half of the spectrum is also
well fit by the model. However, $\chi^2$ can be reduced further
if some A/F stars are included `manually'. One
explanation for this
could be the above mentioned dilution of the stellar continuum in
the nucleus with light from the only partially resolved blob BB3
(which is strong in A/F stars), and/or maybe even smaller, completely
unresolved blobs.

The procedure thus fits the combination of main sequence stars
(O5, B5, A5, F5 to sample the young stellar component, and G5, K5,
and M5 stars to fine-tune the old galaxy population)
and template central cluster galaxy to the main continuum regions
blueward of 4600\thinspace\AA. Wavebands containing emission lines were
excluded from the fit. The code was run iteratively, with all
components included in the first go. The algorithm automatically sets
to zero components that are not required. The fitting procedure was
repeated until no significant improvement of $\chi^2$ could be achieved.

Obviously the results obtained are sensitive to the intrinsic
reddening inferred from the Balmer line ratios (as discussed in
section~\ref{extinction}) and the error bars on the assumed values of
{\sl E(B-V)} are still relatively large. There is the additional
possibility that the young stellar populations causing the blue light
also have an increased Balmer absorption that may affect this
ratio. (Also see Sec. \ref{discussion} for a discussion of the applicability of
{\sl E(B-V)} derived from the properties of the gas to the stellar component.)
Hence we fit both, the uncorrected and corrected spectra, so that
the results on the ones without any de-reddening can be taken as a very
conservative lower limit for the true amount of blue light and number
of early stars present. For BB3 we are only discussing the results from the
fit to the uncorrected data, as it is not possible to achieve a good
fit assuming the value of {\sl E(B-V)}$=0.73$ originally derived from
the Balmer
decrement for this region (see discussion in Sec. \ref{continuum}).

\subsubsection {Stellar synthesis results} 
\label{modelresults}

Although it is possible to obtain apparently acceptable fits to the
nucleus with simply the template central cluster galaxy spectrum,
$\chi^2$ could be reduced further if small amounts of A/F stars were
included. At PA305 there is even the possibility that a marginal O star
component is present. This, and the varying relative importance of the A and F
stars in the nuclear spectra at PA305 and PA258 respectively, agrees
with the earlier suggestion in section \ref{extinction} that the PA258
either suffers from a small amount of atmospheric dispersion or that
it is not centred on exactly the same region as the other PA.

\begin{figure*}
\begin{minipage}{175mm}
\includegraphics[width=70mm,height=85mm,angle=270]{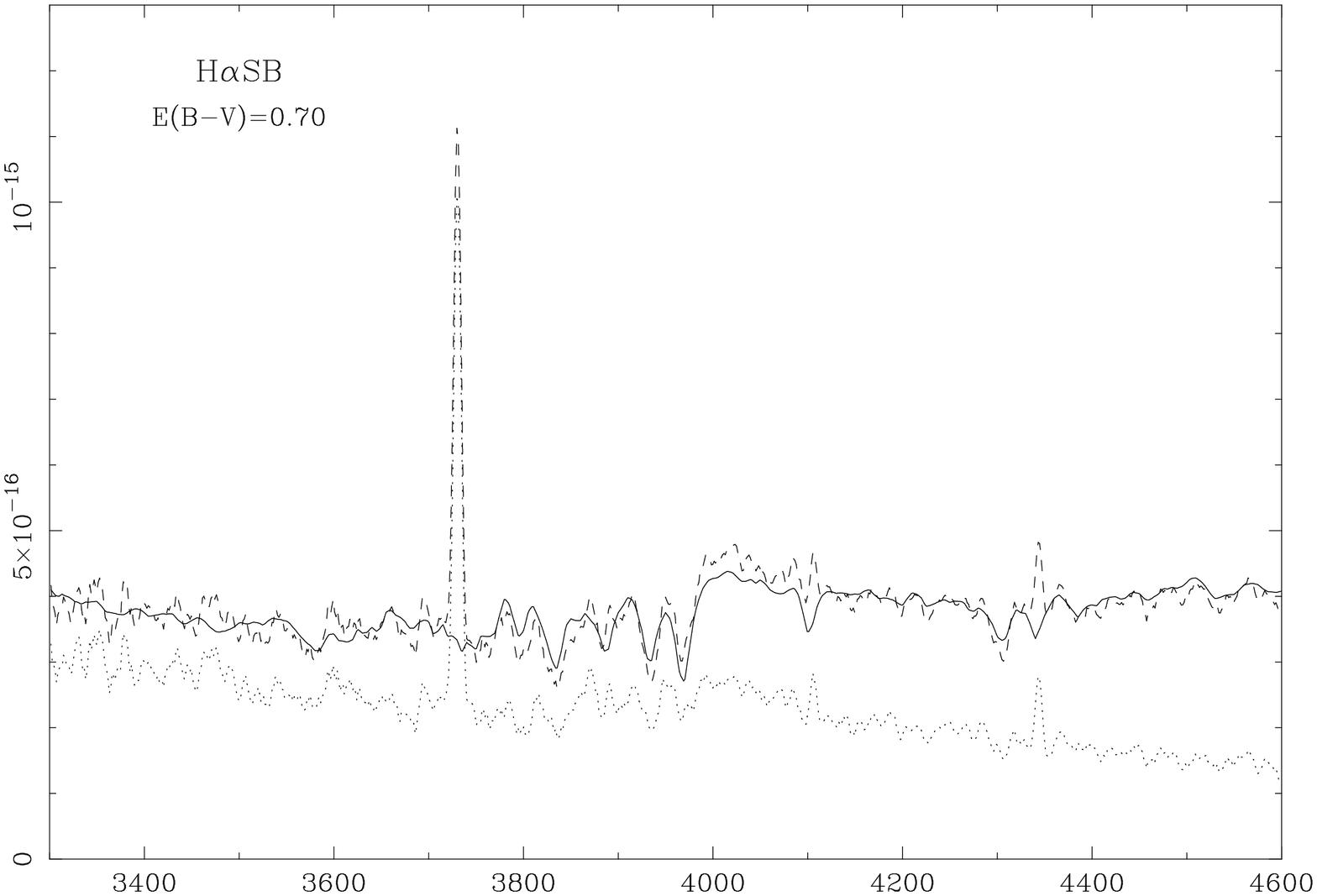}
\includegraphics[width=70mm,height=85mm,angle=270]{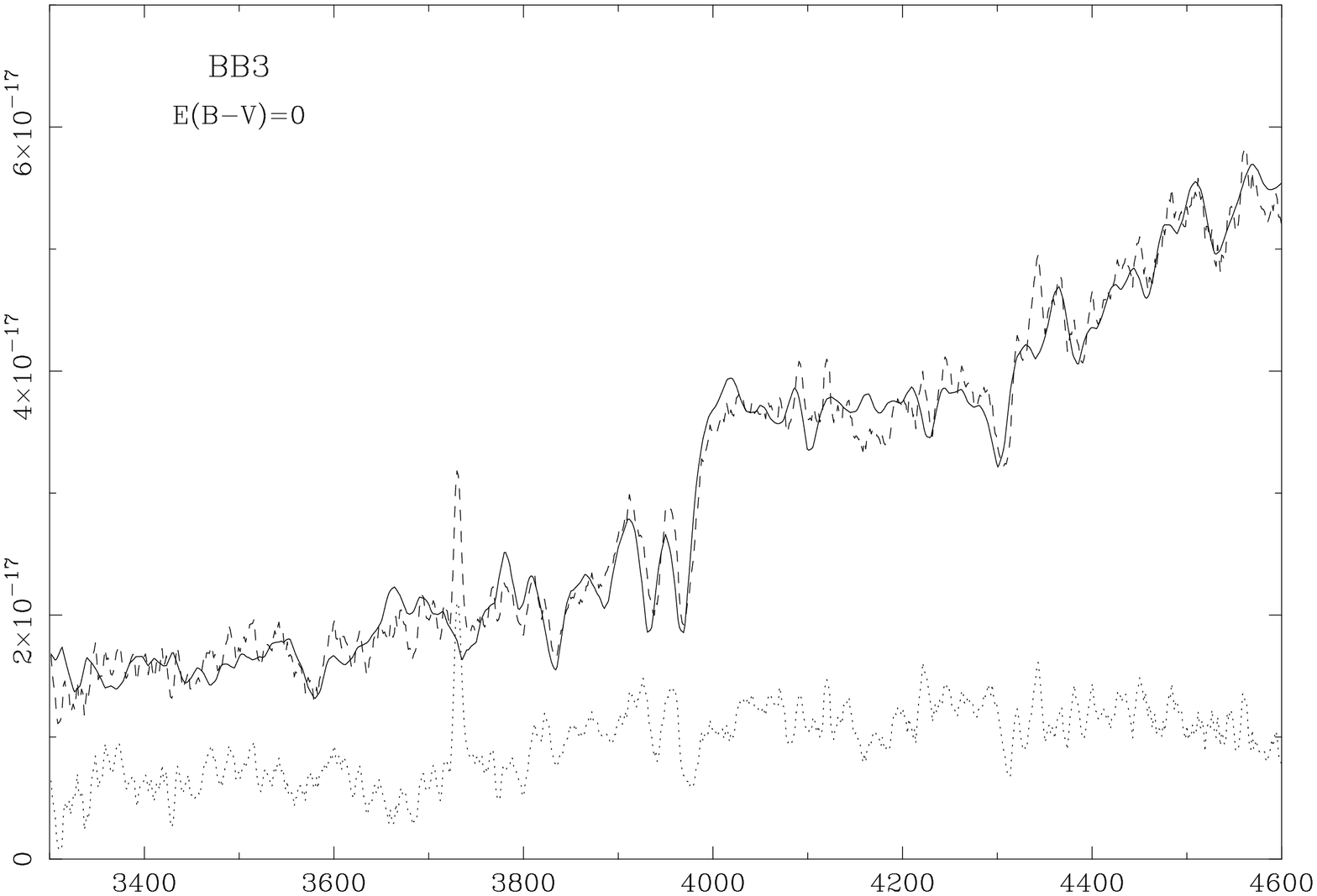}
\newline
\includegraphics[width=70mm,height=85mm,angle=270]{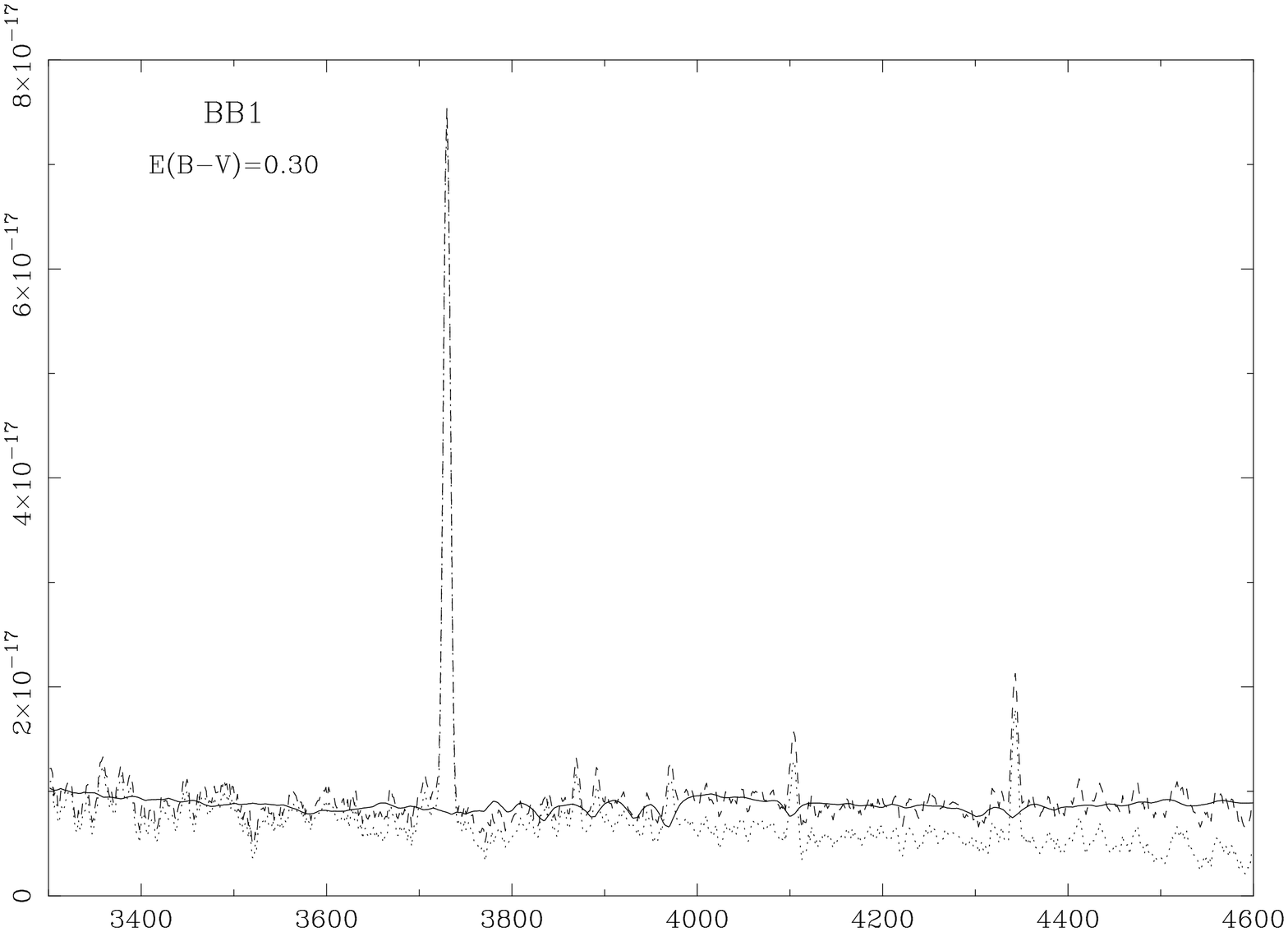}
\includegraphics[width=70mm,height=85mm,angle=270]{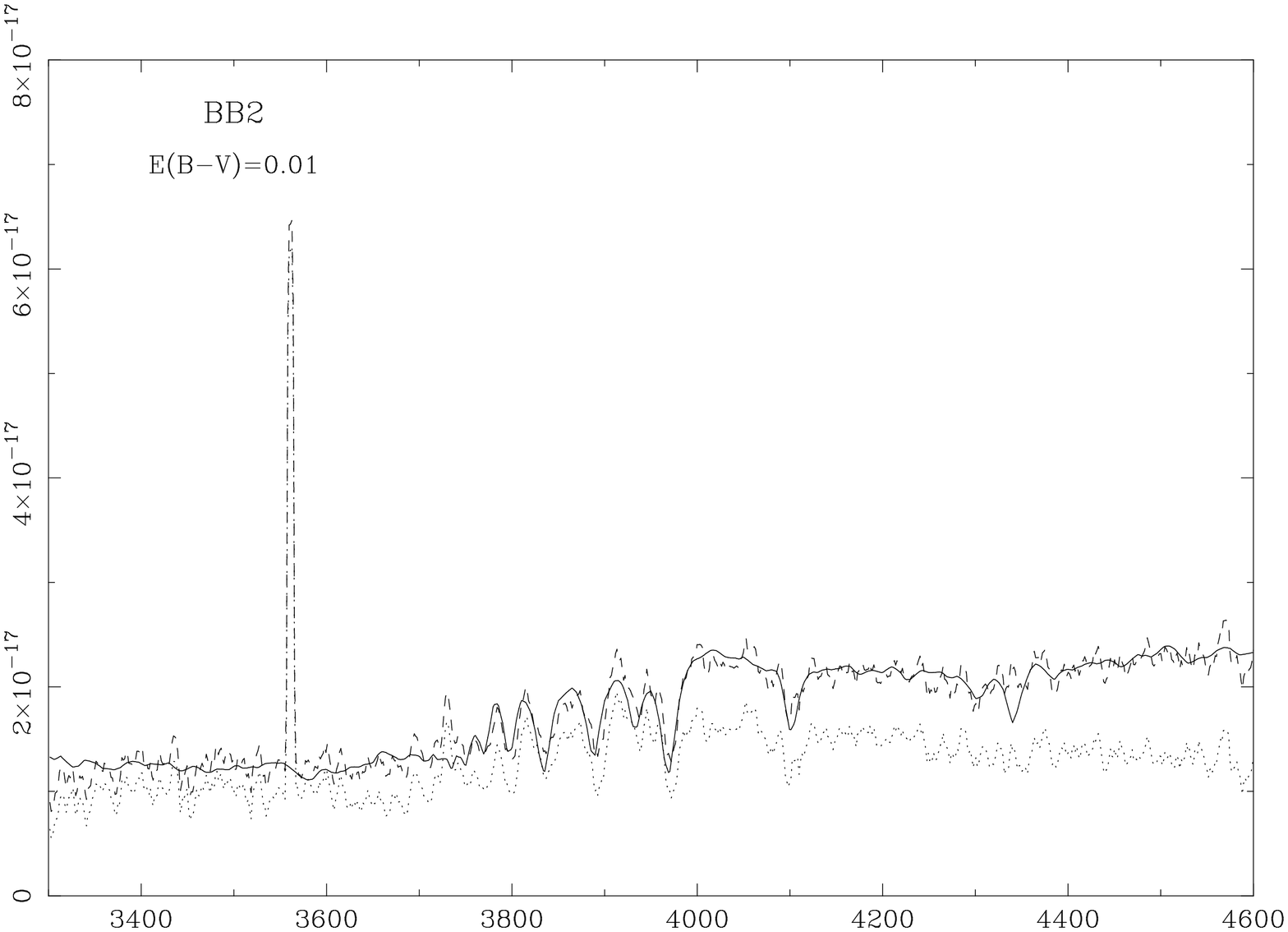}
\end{minipage}
\caption{Fits to the de-reddened spectra of H$\alpha$SB and BB1 (panels
  on LHS), and BB3
and BB2 (panels on RHS). The data are represented as a dashed line
and the fit as a solid line. The dotted line is a difference spectrum
with the late component subtracted,
showing only the early component (O to F stars + emission lines) of the
data. Fluxes are given in \ergpspcmsqpA and any corrections made for
intrinsic reddening are shown in the respective LH corner of the
plot. All spectra are smoothed.}
\label{blobfits}
\end{figure*}

The best fits to the spectra of the different regions in the RX~J0821
system are shown in Fig.~\ref{blobfits}. 
The fit to the H$\alpha$SB continuum is very good, although the slight
under-estimation of
the redwards side of the 4000\thinspace\AA\ break is typical for these
kind of fits (e.g.~figs in Crawford \& Fabian 1993).
The model includes significant proportions of both A- and O-stars added
to the template/late star component spectrum. Note that before
de-reddening, even though there is an additional F-component, still a
significant number of O-stars is required.

BB1 shows quite a different spectrum (Fig. \ref{blobfits}, lower
left), with a remarkably strong Balmer series and
[OII]$\lambda$3727\thinspace\AA{} line in emission. The model uses
mainly O-stars, plus a significant A-star component added on to the
template galaxy, again confirming the presence of a blue excess.
The apparent absence of any B-stars in BB1 and even more obviously in
H$\alpha$SB can be explained by two separate starbursts having
taken place in these blobs. However, it cannot be ruled out that there
is a certain
amount of confusion between O and B stars due to the similarity of
their spectra and the scarcity of distinguishing spectral
features in these stars. (Note that the regions with prominent Balmer
lines had to be excluded from the fit due to the strong emission lines
in the blobs.)

The best fit to BB3 (Fig.~\ref{blobfits}, upper right panel) includes
statistically insignificant proportions of O- and B-stars, as
well as significant contributions of A- and F-stars added on to the late
component indicating an older starburst. Note, however, that we are
using a minimum assumption of
{\sl E(B-V)}$=0$ for this region. Any amount of reddening to be
corrected for would increase the importance of early stars in this fit
significantly.

Finally, the fit to the region BB2 is shown in Fig.~\ref{blobfits} (lower
right panel). The fit reproduces the general shape and strong Balmer
absorption series seen in the spectrum by using a marginally
significant O-star component and significant amounts of all other
stellar components. (The strong emission feature at
$\sim$3560\thinspace\AA{} is a cosmic ray event.)

The difference spectrum (data minus the old component) for every blob
is shown as a dotted
line in this figure. A strong blue excess is obvious for H$\alpha$SB
and BB1. BB2 and
BB3 show a somewhat less pronounced blue light component stemming from
older starbursts than the other two.

Tab. \ref{stellarfluxes} shows the monochromatic slit fluxes at
$\sim4500$\thinspace\AA{} for each
stellar component and the template galaxy used in the respective
model before and after correction for internal reddening. Whenever
the code did not require a stellar component for the final fit, a $1\sigma$
upper limit is given. The slit fluxes have been converted into
luminosities and then into approximate numbers of stars of each type 
(Kurucz~1979; Allen~1995) also given in
Tab. \ref{stellarfluxes}. The errors are the $1\sigma$
($\Delta\chi^2=1$) confidence limits on a single interesting
parameter, derived by manually stepping the individual fluxes in
$\chi^2$ space.

\begin{table*}
\begin{center}
\begin{tabular}{|l|l|c|c|c|c|c|}
\hline
{\sl Region} && {\sl O5} & {\sl B5} & {\sl A5} & {\sl F5} & {\sl Galaxy}\\
\hline
nuc & $E=0$ & {\bf 0.21}$^{+0.17}_{-0.19}$ & $\approxlt 0.35$ & {\bf
0.50}$^{+6.5}_{-0.6}$ & {\bf 3.4}$^{+2.0}_{-2.0}$ & {\bf
31.2}$^{+2.5}_{-1.3}$\smallskip\\
 && ($4.3^{+3.5}_{-3.9}\times 10^2$) & ($\approxlt 7.47\times 10^4$) &
($2.0^{+27.4}_{-2.5}\times 10^6$) & ($1.16^{+0.68}_{-0.68}\times 10^8$) &\\
\hline

H$\alpha$SB & $E=0$ & {\bf 2.18}$^{+0.16}_{-0.18}$ & $\approxlt 0.17$
 & {\bf 2.8}$^{+0.6}_{-0.8}$ & {\bf 3.7}$^{+1.5}_{-1.5}$ & {\bf
 15.1}$^{+1.3}_{-1.4}$\smallskip\\
 && ($4.4^{+0.3}_{-0.4}\times 10^3$) & ($\approxlt 3.73\times 10^4$) &
 ($1.18^{+0.25}_{-0.34}\times 10^7$) & ($1.27^{+0.51}_{-0.51}\times
 10^8$) &\medskip\\

& $E=0.70$ & {\bf 100.3}$^{+1.7}_{-1.8}$ & $\approxlt 2.8$ & {\bf
66.5}$^{+5.5}_{-5.3}$ & $\approxlt 4.7$ & {\bf
 100.6}$^{+21.8}_{-23.9}$\smallskip\\
 && ($2.04^{+0.03}_{-0.04}\times 10^5$) & ($\approxlt 8.14\times
10^5$) & ($2.80^{+0.23}_{-0.22}\times 10^8$) & ($\approxlt 1.61\times
10^8$) &\\
\hline

BB1 & $E=0$ & {\bf 0.60}$^{+0.02}_{-0.02}$ & $\approxlt 0.18$ & {\bf
 0.54}$^{+0.12}_{-0.12}$ & $\approxlt 0.54$ & {\bf
1.44}$^{+0.38}_{-0.39}$\smallskip\\
 && ($1.22^{+0.04}_{-0.04}\times 10^3$) & ($\approxlt 3.95\times
10^4$) & ($2.28^{+0.51}_{-0.51}\times 10^6$) & ($\approxlt 1.85\times
10^7$) &\medskip\\

& $E=0.30$ & {\bf 2.71}$^{+0.08}_{-0.08}$ & $\approxlt 0.56$ & {\bf
1.64}$^{+0.34}_{-0.31}$ & $\approxlt 1.73$ & {\bf
4.60}$^{+0.38}_{-0.42}$\smallskip\\
 && ($5.50^{+0.16}_{-0.16}\times 10^3$) & ($\approxlt 1.23\times
10^5$) & ($6.91^{+1.43}_{-1.31}\times 10^6$) & ($\approxlt 5.93\times
10^7$) &\\

\hline
\hline

nuc & $E=0$ & $\approxlt 0.003$ & $\approxlt 0.01$ & {\bf
2.92}$^{+0.32}_{-0.21}$ & $\approxlt 0.03$ & {\bf
15.09}$^{+0.64}_{-0.99}$\smallskip\\
 && ($\approxlt 6\times 10^0$) & ($\approxlt 2.19\times 10^3$) &
($1.23^{+0.13}_{-0.09}\times 10^7$) & ($\approxlt 1.03\times 10^6$) &\\
\hline

BB3 & $E=0$ & {\bf 0.22}$^{+0.33}_{-0.27}$ & {\bf 0.3}$^{+1.0}_{-1.0}$
 & {\bf 7.5}$^{+1.2}_{-1.1}$ & {\bf 4.6}$^{+0.9}_{-1.9}$ & {\bf
 21.7}$^{+2.1}_{-1.1}$\smallskip\\
 && ($4.5^{+6.7}_{-5.5}\times 10^2$) &
 ($6.6^{+21.9}_{-21.9}\times 10^4$) & ($3.16^{+0.51}_{-0.46}\times 10^7$) &
 ($1.58^{+0.31}_{-0.65}\times 10^8$) &\\
\hline

BB2 & $E=0$ & {\bf 0.43}$^{+0.25}_{-0.40}$ & {\bf
4.80}$^{+1.37}_{-0.80}$ & {\bf 5.24}$^{+1.00}_{-1.79}$ & {\bf
2.12}$^{+0.92}_{-1.25}$ & {\bf 6.53}$^{+1.74}_{-0.93}$\smallskip\\
 && ($8.73^{+5.07}_{-8.12}\times 10^2$) & ($1.05^{+0.30}_{-0.18}\times
10^6$) & ($2.21^{+0.42}_{-0.75}\times 10^7$) &
($7.26^{+3.15}_{-4.28}\times 10^7$) & \medskip\\

& $E=0.01$ & {\bf 0.45}$^{+0.31}_{-0.39}$ & {\bf 5.14}$^{+1.36}_{-0.78}$ & {\bf
5.34}$^{+1.26}_{-1.70}$ & {\bf 1.84}$^{+1.31}_{-2.58}$ & {\bf
6.57}$^{+1.74}_{-1.73}$ \smallskip\\
 && ($9.13^{+6.29}_{-7.92}\times 10^2$) & ($1.13^{+0.30}_{-0.17}\times
10^6$) & ($2.25^{+0.53}_{-0.72}\times 10^7$) &
($6.30^{+4.49}_{-8.84}\times 10^7$) & \\
\hline

\end{tabular}
\caption{Monochromatic fluxes (at 4500\thinspace\AA) for the stellar and galaxy
components. The first three sections are for PA 305\deg, the following
three for PA 258\deg. Results are given with/without correction for
intrinsic reddening. The values in brackets give the approximate
number of stars in each spectral class. All fluxes are quoted in units of
$10^{-18}$\ergpspcmsq. The $1\sigma$ errors are given (upper limits
for those components not required by the fit).}
\label{stellarfluxes}
\end{center}
\end{table*}

If we assume the respective masses $M$ of O5, B5, A5, and F5 stars to be
40.0, 6.5, 2.1, and 1.3\thinspace$M_{\odot}$ (Zombeck 1990) and their typical
lifetimes $t$ to be $3.7\times 10^6$, $3.9\times 10^7$, $6.0\times 10^8$,
and $3.3\times 10^9$\thinspace yr, respectively (Shapiro \& Teukolsky
1983; Allen 1995), we can make a rough estimate of the mass deposition
rate in stars ($\dot{M}_{\rm Star}$)
for each blob using $\dot{M}_{\rm Star}=\frac{N\cdot M}{t}$, where $N$
is the number of stars. If we sum
over the stellar components, we obtain 3.2, 0.1, 0.2, and
0.3\Msunpyr for H$\alpha$SB, BB1, BB3, and BB2, respectively, which
gives a total $\dot{M}_{\rm Star}=3.8$\Msunpyr. In order to correct
this for the mass contained in late type stars, we can make a crude
estimate using Kroupa, Tout \& Gilmore's (1993) IMF

\begin{equation}
\xi(m)=\left\{ \begin{array}{ll}
                0.035m^{-1.3} & 0.08\leq m <0.5\\
                0.019m^{-2.2} & 0.5\leq m <1.0\\
                0.019m^{-2.7} & 1.0\leq m <\infty.\\
               \end{array}
       \right.
\end{equation}

\noindent The total mass of stars $M_T$ associated with a burst of
star formation is then

\begin{equation}
M_T= \int_{M_{\rm low}}^{M_{\rm high}} m\xi(m)\mbox{d}m.
\end{equation}

\noindent Assuming our stellar components as representative for the mass
range of stars of types O -- F, we adopt $M_{\rm low}=1$, $M_{\rm
high}=120$\Msun{} for the early stars used in our model. The late type
stars are assumed to span the range from $M_{\rm low}=0.1$ to $M_{\rm
high}=1$\Msun. By integrating over the IMF we can work out the correction
factor for the late-type stars, and obtain a corrected total
$\dot{M}_{\rm Star}\approx 8.8$\Msunpyr. Hence we are accounting for less
than one third of the total mass deposition
rate within the central 20\kpc{} of the cluster as derived from the
X-rays (see Sec. \ref{cooling}). Of course, the actual
$\dot{M}_{\rm Star}$ will be somewhat higher, considering that the slits
only sample a certain fraction of the object. For this reason, and
because of the large uncertainties associated with the reddening, we refrain
from carrying out a more detailed study, and merely note that as a first
approximation the cooling 
flow provides enough material to be able to ``feed'' the
starbursts. In order to attempt answering the question of precisely what
percentage of
the material deposited by the flow could actually be turned into stars, we
would need spectra covering the entire extent of the
star-forming region, while at the same time being
of sufficiently high quality to give us a much better handle on the
intrinsic reddening. It is quite possible that $\dot{M}_X=\dot{M}_{\rm
Star}$.

\section{Summary of observational results}

The {\sl Chandra} data show that the central galaxy of RX~J0820.9+0752 is
embedded in cluster gas with a central temperature of 1.8\keV and
moderate X-ray
luminosity. Assuming that no heat sources are present, the observed surface
brightness profile implies that the hot
intracluster gas is cooling within a radius of 20\kpc\ at rates of a
few tens of solar masses a year. The X-ray emission is clearly
extended by around 22\kpc\ towards the NW of the central galaxy. The
scale and orientation of this extended emission is well matched to the
region within the break of the $\dot{M}$ profile, the luminous \ha\
nebula, and the strong CO emission.
The line-emitting nebula cannot be seen in the HST image due to the
latter's relatively blue passband. The image shows, if anything, a
sharp decline in the diffuse
continuum in this system in the same region at radii beyond around
8\kpc. Thus it appears the strong extended X-ray/\ha\ feature is
heavily obscured by dust at shorter wavelengths which agrees with our
estimates of {\sl E(B-V)} obtained from the Balmer decrement in this
region. The HST image does,
however, reveal large clumps of emission (blue continuum and/or
blue emission lines) that cross in an arc from this region to the SE of the
central galaxy; a secondary galaxy lies $\sim$11\kpc\ in this
direction, opposite to the X-ray/\ha\ extension.

The radio power of RX~J0821 is the third lowest of all
line-emitting CCGs in the BCS (only A262 and RX~J1733.0+4345/IC1262
have lower powers) and may not be associated with the core
of the CCG (as is found in the vast majority of BCS
line-emitting CCGs) but instead be coincident with the
main star-forming regions. The discovery of
radio emission from star formation rather than nuclear activity 
is intriguing and needs more detailed radio imaging to 
verify.

We have taken spectra from two slit position angles across the main
galaxy, including the extended nebula to the NW and across the two
arcing lines of clumps seen in the HST image to the ENE. We detect \ha\
in emission out to 24\kpc\ to the NW of the galaxy. The gas in this
extended nebula is redshifted by a roughly constant amount of
$\sim100$\kmps\ relative to the central galaxy, and then undergoes a
strong gradient of about 150\kmps in 8\kpc\ as it crosses this galaxy
and out to the SE. The nucleus of the main galaxy shows little line
emission, no intrinsic reddening or evidence for an excess blue
continuum component. The radial velocities of the stellar and nebular
components in the nucleus of the CCG are in very good agreement, but
they are offset significantly by $\sim 200$\kmps to the NW, and this
separation appears to be gradually decreasing to almost zero towards
the E. 

There is a marked overall decline in line-width from about 300\kmps\ to 
just below 200\kmps\ over the extent of the nebula from the NW towards
the nucleus, followed 
by a sharp increase back to 300\kmps\ about 3\kpc\ from the
nucleus. The line-width then decreases again over the inner region of
the galaxy.

We isolate two main regions of excess line and continuum emission to
the NW of the central cluster galaxy. The \ha SB feature lies at the
end of the inner arc of clumped emission seen in the HST image, where
it tails off into the core of the
\ha\ nebula. Consistent with its position also at the edge of the 
sharp `bite' of extinction, the strong line emission shows a Balmer decrement
indicating a large degree of intrinsic reddening, with {\sl
E(B-V)}$\sim0.7$. The excess blue continuum is well fit by a
combination of O and A-type stars. 

A more isolated blue blob (BB1) lies further to the NW, still within
the core of the \ha\ nebula. Its spectrum again shows a strong excess
blue continuum (again well fit by O and A main sequence stars), and a
strong Balmer series in line emission. Again, the Balmer
decrement indicates a degree of intrinsic reddening at {\sl
E(B-V)}$\sim0.3$. Generally reddening is found to be very high within
the region occupied by the \ha\ nebula.

We also obtain the spectra of two continuum blobs, one from each of
the clumped lines of emission to the ENE of the main galaxy. BB3 is a
conspicuous bright blob forming part of the inner arc, that is bright
in \ha, lying within the peak of emission. A high level of intrinsic
reddening is indicated by the strong Balmer decrement ({\sl
E(B-V)}$\sim0.73$) and
evidence for a small amount of excess blue continuum, fit by a range
of main sequence stars. BB2 forms part of the outer crossing arc to
the ENE, and shows little line emission and an insignificant amount of
intrinsic
reddening. Its blue continuum, however, -- otherwise quite similar to
that of BB3 -- is remarkable in showing a
clear Balmer line sequence in absorption that is fit by a complex
mixture of stars of all types.

The emission line gas at the locations of BB1, BB3 and \ha SB has a
lower [NII]/\ha, smaller linewidths and stronger \ha\ emission than
found in the surrounding regions. BB2 deviates in this and other
respects from the other blobs.

\section{Discussion}
\label{discussion}

The immediate environment of the central galaxy of the RX~J0820.9+0752
cluster is very complex. The primary region of cooling X-ray gas
contains a luminous \ha\ nebula, high intrinsic absorption, and a weak
radio source. The total slit \ha\ luminosity of the nebula is around
$1.3\times10^{42}$\ergps; as our slit covers an area of about
35\kpc$^2$ at the object's redshift, this corresponds to a \ha\
surface brightness of $\sim 3.6\times 10^{40}$ $\erg\s^{-1}\kpc^{-2}$.
RX~J0821 is therefore an intermediate \ha\ emitter in terms of both
\ha\ luminosity and surface brightness when compared to the BCS
objects (Crawford \etal 1999; compare also their Fig. 6). The
equivalent hydrogen column density derived from the measured Balmer
decrement is $3.7\pm 0.5\times 10^{21}$\apc (assuming a gas-to-dust
ratio of N(HI)/{\sl E(B-V)}$=4.8\times10^{21}$ atom\thinspace
cm$^{-2}$\thinspace mag$^{-1}$; Bohlin, Savage \& Drake 1978). This
does not seem to agree very well with the value of X-ray determined $N_{\rm
H}=3.2_{-2.2}^{+2.5}\times 10^{20}$\apc{} given in
Tab. \ref{xmodel}. However, given the large region covered by the
X-ray annulus, the uncertainties associated with both the
value for $N_{\rm H}$ used in the X-ray model and that derived from
the Balmer decrement, and especially projection effects, this may not
be significant. Furthermore
our slit was chosen so as to sample the most interesting regions of
the object which are associated with heavily obscuring dust
clouds. Hence the high average {\sl E(B-V)} derived from the slit is
probably not representative for the whole object.


It is interesting that the surrounding cluster hosts a cooling flow,
despite the complicated morphology of the CCG and a head-tail
radio source that could indicate a cluster that is not fully relaxed. The
coincidence between the X-ray and \ha\ emission is reminiscent of the
filament found in A1795 (Fabian \etal 2001a), where a 40-kpc-long X-ray
filament is also associated with a similar region of high optical
reddening and strong line emission. Half-way down this filament lies a
large clump of blue light very similar to the blue blobs we find
around RX~J0821 (Cowie \etal 1983; McNamara \etal 1996). Hu \etal (1985) show
that the outer emission line nebulosity in A1795 is offset in velocity
by an average 150\kmps from the galaxy whereas the line emission at
the position
of the galaxy shares its redshift. The current best explanation for
the observed X-ray/\ha\ filament in A1795 is that it is a cooling wake
prompted to collapse from the intracluster medium by the passage of
the CCG through the cluster core. Fabian \etal (2001a) give other
explanations for such a feature as well, but state that the only
likely mechanism apart from the cooling wake is that the filament is a
contrail produced by the ram pressure of a radio source passing
through a multiphase medium. The similarity of the extended X-ray/\ha\
structure in RX~J0820.9+0752 to that of A1795 strongly suggests that similar
mechanisms are at work here. Given that the radio source in RX~J0820.9+0752
is very weak, its similarity to A1795 strengthens the interpretation
that a cooling wake rather than radio source is responsible for these
features in both galaxies. In addition, the velocity gradient within
the extended ionized gas around RX~J0821 is of the same magnitude
(150\kmps) as in A1795.

We find a velocity
offset between the emission line gas and the stars that diminishes
from about 200\kmps at the position of the \ha SB to the NW to being
practically consistent with each other in the region of the eastern
arcs. This is sampled reliably at only three points (\ha SB, BB3, and
BB2), but the data seem consistent with the velocity offset vanishing
somewhere near the secondary galaxy. Furthermore, the latter's stellar
radial velocity of $v_{rad}=+77\pm 32$\kmps relative to the CCG is
consistent with that of most of the off-nucleus gas. The connection of
the extended X-ray/\ha\ feature to the CCG suggests that it is clearly
produced by the CCG, but we also see a kinematic suggestion for an association
of the gas with the secondary. It is plausible that some kind of
tidal triggering has taken place in the past, contributing to the
formation of the observed arcs with their blue blobs on the Eastern
side of the galaxy. For example, if the second galaxy passed through
the cooling wake left by the CCG, it could have dragged some of the
gas out of the wake in the process.

We can use the break in the mass-deposition profile (Fig.
\ref{fig:xproj}, bottom panel) to estimate the effective age of the cooling
flow. Allen \etal (2001c) identify this break with the outermost radius
at which cooling occurs. In RX~J0820.9+0752 the break occurs at $r\approx 28$\kpc. The
cooling time, and therefore the effective age of the flow, at this
radius is $<1$\thinspace Gyr: the cooling flow is relatively young, which is
additional support for some kind of recent merger activity.

In Tab. \ref{bloblum} a comparison between \ha\ luminosity as observed
and predicted from the ionizing flux of the O-stars derived from the
continuum fits applying the method described in Allen (1995)
both before and after correction for intrinsic reddening is given. The results
for BB1 and BB2 are in good agreement, even though BB2 does allow for
an additional source of ionization to be present. This is even more
true for BB3, where the observed luminosity is significantly higher than the
calculated one. (We note however, that if reddening is underestimated,
this could also account for the observed \ha.) 
In the H$\alpha$SB the calculated luminosity is significantly higher
than the observed one. However, if we consider the result for {\sl
E(B-V)}$=0$, it is the other way around, which could be
an indication that the reddening might have been over-estimated somewhat.

As a caveat, we note that their velocity offset means that the stellar
and gas components of the blobs may not be cospatial. Furthermore, if
most of the dust responsible for the Balmer decrement was associated
with the gas rather than the stars, the {\sl E(B-V)} derived from the
emission lines would over-correct the stellar component. This is one
reason why we have regarded the reddening corrections with great
caution and have included results assuming {\sl E(B-V)}$=0$ throughout
the paper as well.

\begin{table}
\begin{center}
\begin{tabular}{l|c|c}
& $L(\mbox{H}\alpha)_{obs}$ & $L(\mbox{H}\alpha)_{calc}$\\
\hline\hline
H$\alpha$SB & 7/37 & $2.4^{+0.2}_{-0.2}$/$112^{+2}_{-2}$\\
\hline
BB1 & 1.7/3.0 & $0.67^{+0.02}_{-0.02}$/$3.03^{+0.09}_{-0.09}$\\
\hline
BB2 & 1.0/1.0 & $0.48^{+0.28}_{-0.45}$/$0.50^{+0.35}_{-0.44}$\\
\hline
BB3 & 5 & $0.2^{+0.4}_{-0.3}$\\
\hline
\end{tabular}
\caption{Measured and calculated \ha\ luminosity before/after
correction for reddening for each blob. Luminosities are
given in units of $10^{40}$\ergps, and $1\sigma$ errors are given for
the calculations.}
\label{bloblum}
\end{center}
\end{table}

This interpretation would imply that the stars might not
be a viable ionization source for the gas. Therefore it
is worth while to consider other possible sources for the excess \ha\
emission such as the cold mixing
process described in Fabian \etal (2001b). If the material that has cooled to
about 1\keV\ mixes with cold ($T\sim 10^4$K) gas, it then undergoes a
rapid phase of further cooling. The `missing' soft X-ray luminosity
is thus re-radiated at optical/UV wavelengths and could explain the
strength of the emission lines in this object. If we assume that all
the observed slit \ha\ luminosity
is due to re-radiated X-ray emission from cold mixing and convert it
into an equivalent mass deposition rate $\dot{M}_{H\alpha}$ using

\begin{equation}
\dot{M}_{H\alpha}=2
\left(\frac{L_{H\alpha}}{10^{40}\mbox{ergs}^{-1}}\right)
\left(\frac{kT}{1\mbox{keV}}\right)^{-1}\mbox{M}_{\odot}\mbox{yr}^{-1},
\end{equation}

\noindent (Fabian \etal 2002), we obtain $\dot{M}_{H\alpha}=38$\Msunpyr
for the uncorrected slit mass deposition rate. This increases to
$\dot{M}_{H\alpha}=260$\Msunpyr if correction for reddening is
applied. 
Obviously, comparing to the previously derived
$\dot{M}_X\approx 30$\Msunpyr even when including ionization due to
star formation, there is more than enough room for emission due to
cold mixing. 

Taken together, it seems that several processes are at work in the
central region of RX~J0820.9+0752: although there are no redshift
measurements for other potential cluster members that could provide us
with a redshift for the whole cluster and thus we cannot test the
motion of the CCG, it is consistent with our observations
that the CCG moves within the cluster
potential, leaving behind a wake of cooling gas as observed in the X-ray
and \ha\ light. A certain degree of interaction with the secondary
galaxy (be it direct interaction between the two galaxies or the
secondary passing through the cooling wake of the CCG) has helped
produce dense, off-nucleus clumps of gas in which
star formation is triggered as apparent from their blue continua. 
The observed strong \ha\ emission could be explained by either cold
mixing or (if stars and gas are cospatial) stellar ionization. Most likely both
processes have made a contribution to produce the observed line luminosity.

It is clear further features similar to our `blue blobs' exist around
RX~J0821, not only in the two HST arcs of emission, but there also
appears to be another isolated bright blob at the edge of the absorbed
region, around 5.5 \arc{} to the WNW of the main galaxy. Studying more
of these blobs in detail should allow us to learn more about the
intricate interplay of a massive galaxy moving through the ICM of a
cooling flow cluster, the associated cooling wake, possible tidal
interaction with the secondary galaxy, the role of the weak radio
source, and strong star formation activity triggered by one or more of
these processes.

\section{Acknowledgments}
CMB-K thanks the Austrian Bundesministerium f\"ur Bildung,
Wissenschaft und Kunst and the
Cambridge European Trust for financial support, and Robert Schmidt,
Dougal Mackey and Eduardo Delgado Donate for
valuable discussions and general help. ACF, ACE, CSC
and SWA thank the Royal Society
for financial support. We thank Richard Wilman for his assistance with
collection of the data during the WHT observing run, and Drs Frayer and
Ivison for knowledge of their SCUBA and OVRO results in advance of
publication. For the preparation of this paper we made use of data
acquired with
Chandra (NASA-MSFC, Huntsville, Alabama), HST (Space Telescope Science
Institute, Baltimore, USA),
AAT (Anglo-Australian Observatory, Siding Spring, Australia), and WHT
(Observatorio del Roque de los Muchachos, La Palma, Spain). We
acknowledge the use of STScI's Digital Sky Survey (DSS).


\begin{thebibliography}{}
\bibitem{} Allen S.W., 1995, MNRAS, 276, 947
\bibitem{} Allen S.W., Ettori S., Fabian A.C., 2001a, MNRAS, 324, 877
\bibitem{} Allen S.W., Schmidt R.W., Fabian A.C., 2001b, MNRAS, 328, L37
\bibitem{} Allen S.W., Fabian A.C., Johnstone R.M., Arnaud K.A.,
Nulsen P.E.J., 2001c, MNRAS, 322, 589
\bibitem{} Anders E. \& Grevesse N., 1989, GeCoA, 53, 197
\bibitem{} Arnaud K.A., 1996, in Astronomical Data Analysis Software
and Systems V, eds. Jacoby G. and Barnes J., ASP Conf. Series volume
101, p17
\bibitem{} Balucinska-Church M., McCammon D., 1992, ApJ, 400, 699
\bibitem{} Becker R.H., White R.L., Helfand D.J., 1995, ApJ, 450, 559
\bibitem{} Bland-Hawthorn J., Jones D.H., 1998,
Publ. Astron. Soc. Aust., 15, 44
\bibitem{} Bohlin R.C., Savage B.D., Drake J.F., 1978, ApJ, 224, 132
\bibitem{} Br\"uggen M., Kaiser C.R., 2001, MNRAS, 325, 676
\bibitem{} Churazov E., Br\"uggen M., Kaiser C.R., B\"ohringer H.,
Forman W., 2001, ApJ, 554, 261
\bibitem{} Churazov E., Sunyaev R., Forman W., B\"ohringer H., 2002,
  MNRAS, submitted (astro-ph/0201125)
\bibitem{} Cowie L.L., Hu E.M., Jenkins E.B., York D.G., 1983, ApJ,
272, 29
\bibitem{} Crawford C.S., Fabian A.C., 1993, MNRAS, 265, 431
\bibitem{} Crawford C.S., Edge A.C., Fabian A.C., Allen S.W.,
B\"ohringer H., Ebeling H., McMahon R.G., Voges W., 1995,  MNRAS, 274,
75
\bibitem{} Crawford C.S., Allen S.W., Ebeling H., Edge A.C., Fabian
A.C., 1999, MNRAS, 306, 857 (BCS~III)
\bibitem{} David L.P., Nulsen P.E.J., McNamara B.R, Forman W., Jones C.,
Ponman T., Robertson B., Wise M., 2001, ApJ, 557, 546
\bibitem{} Ebeling H., Edge A.C., Allen S.W., Crawford C.S. Fabian
A.C., Huchra J.P., 2000, MNRAS, 318, 333
\bibitem{} Ebeling H., Rangarajan F.V.N., White D.A., 2002, MNRAS, submitted
\bibitem{} Edge A.C., 2001, MNRAS, 328, 762
\bibitem{} Edge A.C., Ivison R.J., Smail I., Blain A.W., Kneib J.-P.,
1999, MNRAS, 306, 599
\bibitem{} Edge A.C., Wilman R.W., Johnstone R.M., Crawford C.S.,
Fabian A.C., Allen S.W., 2002, MNRAS, submitted
\bibitem{} Ettori S.E., Fabian A.C., Allen S.W., Johnstone R.M., 2001,
MNRAS in press 
\bibitem{} Fabian A.C., 1994, A\&AR, 32, 277
\bibitem{} Fabian A.C., Sanders J.S., Ettori S., Taylor G.B., Allen
S.W., Crawford C.S., Iwasawa K., Johnstone R.M., Ogle P.M., 2000,
MNRAS, 318, 65
\bibitem{} Fabian A.C., Sanders J.S., Ettori S., Taylor G.B., Allen S.W.,
Crawford C.S., Iwasawa K., Johnstone R.M., 2001a, MNRAS, 321, L33
\bibitem{} Fabian A.C., Mushotzky R.F., Nulsen P.E.J., Peterson J.R.,
2001b, MNRAS, 321, L20
\bibitem{} Fabian A.C., Allen S.W., Crawford C.S., Johnstone R.M,
Morris R.G., Sander J.S., Schmidt R.W., 2002, MNRAS, 332, L50
\bibitem{} Heckman T.M., Baum S.A., van Breugel W.J.M., McCarthy P.,
1989, ApJ, 338, 48
\bibitem{} Hu E.M., Cowie L.L., Wang Z., 1985, ApJS, 59, 447
\bibitem{} Johnstone R.M., Allen S.W., Fabian A.C., Sander J.S., 2002,
astro-ph/0202071
\bibitem{} Kaastra J.S., Mewe R., 1993, Legacy, 3, 16, HEASARC, NASA
\bibitem{} Kaastra J.S. \etal, 2001, A\&A, 365, L99 
\bibitem{} Kriss G.A., 1994, PASP Conf. Ser., 61, 437
\bibitem{} Kroupa P., Tout C.A., Gilmore G., 1993, MNRAS, 262, 545
\bibitem{} Kurucz R.L., 1979, ApJS, 40, 1
\bibitem{} Liedhal D.A., Osterheld A.L., Goldstein W.H., 1995, ApJ, 438, L115 
\bibitem{} McNamara B.R. \etal, 2000, ApJ, 534, L135
\bibitem{} McNamara B.R., Wise M., Sarazin C.L., Jannuzi B.T., Elston
R., 1996, ApJ, 466, L9
\bibitem{} Peterson J.R \etal 2001, A\&A, 265, L104
\bibitem{} Pickles A.J., 1998, PASP, 110, 863
\bibitem{} Reynolds C.S., Heinz S., Begelman M.C., 2002, MNRAS, 332, 271
\bibitem{} Ruszkowski M., Begelman M.C., 2002, ApJ, submitted
  (astro-ph/0207471)
\bibitem{} Schmidt R.W., Allen S.W., Fabian A.C., 2001, MNRAS, 327, 1057
\bibitem{} Shapiro S.L., Teukolsky S.A., 1983, Black Holes, White
Dwarfs and Neutron Stars. John Wiley \& Sons, New York
\bibitem{} Stark A.A., Gammie C.F., Wilson R.W., Bally J., Linke R.A.,
Heiles C., Hurwitz M., 1992, ApJS, 79, 77
\bibitem{} Tennant A.F., 1991, NASA Technical Memorandum, 4301
\bibitem{} Voigt L.M., Schmidt R.W., Fabian A.C., Allen S.W.,
Johnstone R.M., 2002, MNRAS, submitted (astro-ph/0203312)
\bibitem{} White D.A., Jones C., Forman W., 1997, MNRAS, 292, 419
\bibitem{} Young A.J., Wilson A.S., Mundell C.G., 2002, ApJ, submitted
(astro-ph/0202504)
\bibitem{} Yun M.S., Reddy N.A., Condon J.J., 2001, ApJ, 554, 803
\bibitem{} Zombeck M.V., 1990, Handbook of Space Astronomy \&
Astrophysics, 2nd edn., Cambridge Univ. Press, Cambridge
\end{thebibliography}
\end{document}